\documentclass[acmsmall,screen]{acmart}

\AtBeginDocument{%
  \providecommand\BibTeX{{%
    \normalfont B\kern-0.5em{\scshape i\kern-0.25em b}\kern-0.8em\TeX}}}

\usepackage{array}
\usepackage{subfigure}
\usepackage{graphicx}
\usepackage[switch]{lineno}
\usepackage{amsmath}

\usepackage{amssymb}
\usepackage{multirow}
\usepackage{booktabs}

\usepackage{algorithm}
\usepackage{algorithmic}
\usepackage{xcolor}
\usepackage{xspace}
\usepackage{flushend}
\usepackage{fancyhdr}
\usepackage{makecell}

\newcommand{\paratitle}[1]{\vspace{1.5ex}\noindent\textbf{#1}}
\newcommand{\ie}{\emph{i.e.,}\xspace}

\newcommand{\eg}{\emph{e.g.,}\xspace}
\newcommand{\baby}{{BiPNet}\xspace}
\newcommand{\babyuni}{{BiPNet}}
\newcommand{\etal}{\emph{et al.}\xspace}
\newcommand{\fig}{Figure \xspace}

\setcopyright{acmcopyright}
\copyrightyear{2023}
\acmYear{2023}
\acmDOI{XXXXXXX.XXXXXXX}

\acmJournal{TOIS}
\acmVolume{0}
\acmNumber{0}
\acmArticle{0}
\acmMonth{0}

\begin{document}

\title{Bi-Preference Learning Heterogeneous Hypergraph Networks for Session-based Recommendation}

\author{Xiaokun Zhang}
\email{dawnkun1993@gmail.com}
\affiliation{%
  \institution{Dalian University of Technology}
  \streetaddress{No.2 Linggong Road}
  \city{Dalian}
  \country{China}
  \postcode{116024}
}

\author{Bo Xu}
\email{xubo@dlut.edu.cn}
\affiliation{%
  \institution{Dalian University of Technology}
  \streetaddress{No.2 Linggong Road}
  \city{Dalian}
  \country{China}
  \postcode{116024}
}

\author{Fenglong Ma}
\email{fenglong@psu.edu}
\affiliation{%
  \institution{Pennsylvania State University}
  \streetaddress{University Park, PA 16801}
  \city{Pennsylvania}
  \country{USA}
  \postcode{19019}
}

\author{Chenliang Li}
\email{cllee@whu.edu.cn}
\affiliation{%
  \institution{Wuhan University}
  \streetaddress{Bayi Rd}
  \city{Wuhan}
  \country{China}
  \postcode{430075}
}

\author{Yuan Lin}
\email{zhlin@dlut.edu.cn}
\affiliation{%
  \institution{Dalian University of Technology}
  \streetaddress{No.2 Linggong Road}
  \city{Dalian}
  \country{China}
  \postcode{116024}
}

\author{Hongfei Lin}
\email{hflin@dlut.edu.cn}
\affiliation{%
  \institution{Dalian University of Technology}
  \streetaddress{No.2 Linggong Road}
  \city{Dalian}
  \country{China}
  \postcode{116024}
}

\authornote{Corresponding author}

\newcommand \footnoteONLYtext[1]
{
	\let \mybackup \thefootnote
	\let \thefootnote \relax
	\footnotetext{#1}
	\let \thefootnote \mybackup
	\let \mybackup \imareallyundefinedcommand
}
\footnoteONLYtext{
\noindent This work is supported by the Natural Science Foundation of China (No.62376051, No.62076046 and No.62006034).

\noindent This article is extended from the paper\cite{CoHHN} appearing in SIGIR 2022. In this extension, we (1) further introduce item brand into user behaviors modeling; (2) reconstruct the heterogeneous hypergraph to encode rich information; (3) propose a novel triple-level convolution to aggregate information on the heterogeneous hypergraph, where the co-occurrence, intra-type and inter-type relations among various nodes are explored simultaneously; (4) explore the mutual relations between price and interest preference with the help of multi-task learning; (5) reveal user intent via predicting price and interest preference collectively; (6) conduct data analysis on real-world datasets to support our motivation; (7) add new datasets and up-to-date baselines to verify the effectiveness of the proposed model; (8) conduct a more detailed analysis about the approach, obtain more insights and review more related literature.

}

\renewcommand{\shortauthors}{XK. Zhang et al.}


\begin{abstract}
\hrulefill

\noindent Session-based recommendation intends to predict next purchased items based on anonymous behavior sequences. Numerous economic studies have revealed that item price is a key factor influencing user purchase decisions. Unfortunately, existing methods for session-based recommendation only aim at capturing user interest preference, while ignoring user price preference. Actually, there are primarily two challenges preventing us from accessing price preference. Firstly, the price preference is highly associated to various item features (\ie category and brand), which asks us to mine price preference from heterogeneous information. Secondly, price preference and interest preference are interdependent and collectively determine user choice, necessitating that we jointly consider both price and interest preference for intent modeling.
To handle above challenges, we propose a novel approach Bi-Preference Learning Heterogeneous Hypergraph Networks (\baby) for session-based recommendation. Specifically, the customized heterogeneous hypergraph networks with a triple-level convolution are devised to capture user price and interest preference from heterogeneous features of items. Besides, we develop a Bi-Preference Learning schema to explore mutual relations between price and interest preference and collectively learn these two preferences under the multi-task learning architecture. Extensive experiments on multiple public datasets confirm the superiority of \baby over competitive baselines. Additional research also supports the notion that the price is crucial for the task.
\end{abstract}

\ccsdesc[500]{Information systems~Recommender systems}

\keywords{Session-based Recommendation, Price and Interest Preference, Heterogeneous Hypergraph, Multi-task Learning.}

\received{1 April 2023}
\received[revised]{1 Aug 2023}
\received[accepted]{24 Oct 2023}

\maketitle

\section{introduction}\label{sec:introduction}

Recommender system (RS), which is ubiquitous in the modern information age, becomes a vital tool to alleviate information overload. RS facilitates online consumption by providing personalized services for individuals, especially in the e-commerce websites. Conventional RS~\cite{IKNN,Koren@KDD2008} utilizes user profiles and long-term behaviors to predict their future actions by assuming that the user identification is available. However, with increasingly strict privacy policy and widespread unregistered users, what the RS can access is merely the short behavior sequence of an anonymous user (\ie session), where these existing conventional RS models are no longer applicable. To improve user experience in this situation, session-based recommendation (SBR) is proposed to provide efficient recommendation for anonymous users within a limited number of interactions~\cite{GRU4Rec, NARM, SR-GNN}.

Due to its highly practical value, SBR has received widespread attention since it appeared. Relying on the powerful learning capacity of neural networks, recent methods for SBR model user behaviors via various kinds of neural structure, such as Recurrent Neural networks (RNN)~\cite{GRU4Rec, NARM}, Convolutional Neural Networks (CNN)~\cite{Tuan@RecSys2017, Caser}, Attention Networks (AN)~\cite{SASRec, STAMP, BERT4Rec} and Graph Neural Networks (GNN)~\cite{SR-GNN, LESSR}. Although remarkable performance has been achieved, the existing SBR methods are only committed to modeling the user \textbf{interest preference} in terms of how much a user likes an item. Another important factor, \ie the user \textbf{price preference} in terms of how much money a user would like to pay for the item, is completely ignored by existing methods. Different from other item features (\eg size, color and style) indirectly influencing user bahaviors, it has been validated that the price can significantly determine whether the user would make a purchase~\cite{degeratu2000consumer, umberto2015price}. Moreover, it is purchased items instead of favorite ones that RS should predict, because that only purchase behaviors can bring profits to businesses and make users further interact with the system (like reviews). Therefore, it is urgent that considering price factor in SBR for accurately predicting the user's behaviors. However, there are mainly two challenges to be solved when we aim to incorporate price into SBR.

Firstly, user price preference is diverse and heavily depends on some features of items, \ie category and brand~\cite{umberto2015price, Zheng@ICDE2020}. Taking a common example, a user may buy \emph{expensive camera} for work, while she could also purchase \emph{cheap sellotape} to use in daily life. This example suggests that user price preference can dramatically change based on categories of items. Besides, some users are willing to spend extra money on items of famous brands, which is called \emph{brand effect} in economics~\cite{degeratu2000consumer}. That means item brands can also influence user price preference. Obviously, the relevant categories and brands should be considered when we model user price preference in SBR. Under such situation, various kinds of information will be necessary to characterize user behaviors, such as \emph{a series of items}, \emph{item prices}, \emph{item categories} and \emph{item brands}. Those \textbf{heterogeneous information} brings intractable challenges for the existing methods such as~\cite{GRU4Rec, NARM, BERT4Rec, SR-GNN,LESSR,DHCN,Xia@CIKM2021}, because that all of them are developed to model the single type information (\ie item id). Thus, to incorporate price into SBR, the first challenge is how to copy with such heterogeneous information.

Secondly, user price preference and interest preference are interdependent and collectively determine user choice. It is quite common that users make price-interest trade-offs when shopping. Also, the \emph{price elasticity} is used by economists to describe the phenomenon that the money a user would like to spend on an item fluctuates with her interest in it~\cite{brynjolfsson2000price}. It indicates that price and interest preference influence each other and contribute to user choices together. Taking an example, it is possible that a user buys an expensive item because of strong interest, even though the item price is higher than what she anticipated. In this instance, the user alters her price preference because of the influence of interest. Likewise, a user will also buy items with little interest out of their low price (it is just what frequently happened during the promotions). Obviously, to predict user purchase decisions, we have to explore the \textbf{mutual relations} between price and interest preference and \textbf{jointly model these two preferences}. However, the existing SBR methods~\cite{GRU4Rec, NARM, BERT4Rec, SR-GNN,LESSR,DHCN,Xia@CIKM2021} only use item ID as the supervised signal to model user interest preference, while can neither explicitly capture user price preference nor adequately handle rich dependencies between price preference and interest preference in determining user choice. Therefore, another challenge facing us is how to accurately model purchase intent of users under these two preferences.

For the first challenge, it is intuitive that using general heterogeneous graph to handle \emph{heterogeneous information}~\cite{Hu@EMNLP2019, Pang@WSDM2022}. However, the existing heterogeneous graph-based approaches can only model pair-wise (dyadic) associations~\cite{Sun@WSDM2021}, \eg the relations between (\emph{category}, \emph{price}) or (\emph{brand}, \emph{price}). In our settings, there are complex high-order relations among different nodes, \eg the triadic associations in $<$\emph{price}, \emph{category}, \emph{brand}$>$ jointly affect an item price semantics. Thus, the current heterogeneous graph-based methods are ineffective in our settings. Moreover, constrained by over-smoothing problem, heterogeneous graph-based methods are unable to model the dependencies between distant nodes within a session~\cite{LESSR}. Fortunately, the hypergraph whose hyperedge can connect multiple nodes is devised recently to capture high-order data relations with arbitrary distance~\cite{Zhou@NIPS2006,DHCN}. Adopting the merits of heterogeneous graph and hypergraph, we propose \textbf{heterogeneous hypergraph networks}, which is able to capture complex high-order dependencies among various nodes, to copy with heterogeneous information in SBR.

Specifically, the heterogeneous information we consider in the heterogeneous hypergraph includes item ID, item price, item category and item brand which are directly relevant to user price and interest preference. Besides, three types of hyperedges, \ie feature hyperedge, price hyperedge and session hyperedge, are designed to facilitate price and interest preference learning in the heterogeneous hypergraph. Based on the constructed heterogeneous hypergraph, we devise a triple-level convolution to learn node embeddings, where three kinds of relations among nodes are emphasized, \ie co-occurrence, intra-type and inter-type relations. After that, the price and interest preferences of users are obtained by attention layers. 

As to the second challenge, we should explore the \emph{mutual relations} between price and interest preference and \emph{take both them into account} to predict user behaviors. The multi-task learning schema is proposed to handle multiple relevant tasks jointly, where the mutual relations among these tasks can be leveraged to improve model capability on all tasks at once~\cite{zhang@TKDE2021}. Inspired by this, we propose a \textbf{bi-preference learning schema}. To be specific, besides using item ID to guide the model to learn user interest preference, the bi-preference learning schema also utilizes the price of target item as another supervised signal to formulate user price preference. Under the schema of multi-task learning, it can explore the mutual relations between price and interest preference and predict both of them simultaneously. Eventually, we reveal user purchase intent based on the learned price preference and interest preference.

All in all, to incorporate price into SBR, we propose a novel approach termed \underline{Bi}-\underline{P}reference Learning Heterogeneous Hypergraph \underline{Net}works (\baby), where the main contributions can be summarized as follows:
\begin{itemize}
    \item We identify and emphasize the importance of the price in influencing user behaviors and consider both price and interest preference in SBR to offer personalized services. As far as we know, this is the pioneer work to incorporate price into SBR.
    \item We propose a novel \baby to improve SBR. In \baby, the customized heterogeneous hypergraph networks are devised to handle heterogeneous information for price and interest preference learning. We also develop a bi-preference learning schema to predict price and interest preference simultaneously via exploring rich mutual relations between them. 
    \item We conduct extensive experiments over multiple real-world datasets and the results show that our proposed \baby outperforms existing state-of-the-art SBR approaches. Further analysis also validates the significance of the price for SBR.
\end{itemize}

The rest of the paper is organized as follows. Section~\ref{sec:relate} briefly reviews the literature most relevant to our work. We then conduct data analysis on a real-world dataset to examine the factors influencing user price preference and formulate the task in Section~\ref{sec:motivation}. In Section~\ref{sec:method}, we elaborate the proposed model \baby. The detailed experimental settings are introduced in Section~\ref{sec:experimentalsetting}. Section~\ref{sec:experiments} presents the experimental results. Lastly, in Section~\ref{sec:conclusion}, we conclude the paper and look forward to the future work.

\section{Related Literature}\label{sec:relate}

Considering that this work aims to improve SBR by introducing price, proposing a customized heterogeneous hypergraph and drawing support from multi-task learning, we briefly review the related work from following four aspects: session-based recommendation, price-aware recommendation, heterogeneous graph/hypergraph and multi-task learning.

\subsection{Session-based Recommendation}
With the ability to predict user actions via limited information, session-based recommendation has become a research hot-spot recently~\cite{ludewig@UMUAI2020, Wang@CS2022}. 
Traditional methods use Markov Chain~\cite{MDP} or Matrix Factorization~\cite{BPR-MF} to mine sequential dependencies in user behavior sequences. Enlightened by the Nearest-neighbor-based approaches~\cite{Koren@KDD2008}, some methods calculate the item similarity  (Item-KNN~\cite{IKNN}) or session similarity (SKNN~\cite{Jannach@RecSys2017}) to determine the recommended items. 

As the pioneer work introducing neural networks into SBR, GRU4Rec~\cite{GRU4Rec} utilizes Recurrent Neural Networks (RNN) to tackle the task. Afterwards, many works improve SBR by applying RNN or its variants such as~\cite{HidasiK@CIKM2018, RepeatNet}. NARM~\cite{NARM} introduces attention mechanism into the GRU4Rec to emphasize the user main purpose. CSRM~\cite{CSRM} further boosts the performance of NARM by exploring collaborative information from neighboring sessions.
As another typical neural structure, Convolutional Neural Networks (CNN) are used in SBR to capture user intent like~\cite{Tuan@RecSys2017} with 3D CNN, and Caser~\cite{Caser} with the standard 2D CNN.
Attention networks are also popular in SBR because of its capacity for filtering out irrelevant items~\cite{SASRec, DPAN, DIDN}. Liu \etal \cite{STAMP} proposed a short-term attention/memory priority model (STAMP) to capture both user general interest and current intents. Adopting the cloze objective from Natural Language Processing (NLP), BERT4Rec~\cite{BERT4Rec} applies deep bi-directional self-attention to mine sequential patterns in the task. 
With the ability to capture pair-wise patterns of item transitions, graph neural networks (GNN) have been widely used in SBR recently~\cite{Qiu@TOIS2020, GCE-GNN, Pang@WSDM2022}. As the first method applying GNN in SBR, SR-GNN~\cite{SR-GNN} views each session as a graph and adopts gated graph convolution to obtain item embeddings. LESSR~\cite{LESSR} handles two information loss issues of GNN for SBR, \ie lossy session encoding and ineffective long-range dependency capturing. PosRec~\cite{Qiu@TOIS2021} exploits positional information to improve GNN-based SBR.
Relying on the ability of hypergraph in capturing complex high-order relations among nodes, Wang \etal~\cite{Wang@SDM2021} and Xia \etal~\cite{DHCN} improved SBR by adopting hypergraph architecture. 
To alleviate data sparsity problem, data augmentation technique is also used to enhance model performance such as co-training~\cite{Xia@CIKM2021}, coupled framework~\cite{Wei@WWW2022}, random walks~\cite{Choi@WSDM2022} and graph augmentation~\cite{Chen@SIGIR2023, Yang@WWW2023,Su@WWW2023}. Moreover, some methods aim to capture the fine-grained preference evolution of users via continuum model~\cite{Guo@CIKM2022} or learning multi-granularity
user intent~\cite{Guo@WSDM2022, Zhang@WSDM2023}. 
Considering that existing architectures of neural network perform differently on different context, some works automatically ensemble distinct networks to adapt to various session scenarios~\cite{Chen@SIGIR2022,Cheng@WWW2022}.   
More recently, some methods try to incorporate auxiliary features of item to model user intent from multiple perspectives like item category~\cite{MGS} and item text~\cite{Hou@KDD2022, MMSBR}.
Our earlier work CoHHN~\cite{CoHHN} emphasizes the importance of price factor on influencing user purchase behaviors. In this work, we further extend CoHHN by incorporating item brand, reconstructing the customized heterogeneous hypergraph and proposing a new triple-level convolution to aggregate rich heterogeneous information. Besides, \baby mines mutual relations between price and interest preference with the help of multi-task learning, where both of item ID and price are used as supervised signals to model user behaviors.

\subsection{Price-aware Recommendation} 

It is obvious that price is a key factor considered by users when they are shopping. Surprisingly, little effort has been made to incorporate it into the recommender system for performance improvements. As pioneers considering price in recommender system, Schafer \etal \cite{schafer1999price} explored user price sensitivity to promote online consumption. 
Kamishima \etal \cite{RecSys2011price} integrated price personalization into recommendation algorithms to improve its commercial viability. 
Wan \etal \cite{Wan@WWW2017price} introduced consumer theories from economics into the lager-scale recommender systems for improving the user satisfaction.
Guo \etal \cite{FGCS2018price} studied the impact of multi-category inter purchase time and price on user purchase behaviors. 
Zheng \etal \cite{Zheng@ICDE2020} modeled user price awareness by designing a unified heterogeneous graph where item price and categories are considered. 
Wu \etal~\cite{Wu@ICDE2022} emphasized the item price competitiveness, which is used to measure the advantage of the item’s price over its comparison prices, in predicting user purchase behaviors.
Unfortunately, simply regarding the price factor as an auxiliary information, most of above methods ignore the vital role of price in user purchase decision. Besides, all of them use the item ID to guide the model learning, while failing to explicitly extract user price preference. In addition, there is no efforts bridging price and SBR, and we are the first to fill this gap.

\subsection{Heterogeneous Graph and Hypergraph}
\paratitle{Heterogeneous graph}, also known as heterogeneous information networks, can effectively encode heterogeneous information among multi-type nodes. Many researchers formulate available data as heterogeneous graph and solve corresponding tasks via extracting relevant semantics on the heterogeneous graph. For example, Hu \etal \cite{Hu@EMNLP2019} aggregated node-level and type-level information in heterogeneous graph which consists of short text, topics and entities for short text classification. Pang \etal~\cite{Pang@WSDM2022} built a heterogeneous global graph containing information of historical sessions, item transitions and co-occurrence to capture user preference in SBR. Fan \etal \cite{Fan@KDD2019} offered intent recommendation via propagating heterogeneous information with a metapath-guided method.

\paratitle{Hypergraph} is a graph structure where a hyperedge can connect with more than two nodes~\cite{Zhou@NIPS2006}. Since its hyperedge can contain arbitrary number of nodes, the hypergraph is able to capture high-order relations among distinct nodes~\cite{Wang@SDM2021, DHCN, Sun@WSDM2021}. With this special capacity, the hypergraph has obtained increasing attention recently. For instance, Feng \etal ~\cite{Feng@AAAI2019} presented a hypergraph neural networks (HGNN) framework for learning data embeddings. Jiang \etal ~\cite{Jiang@IJCAI2019} utilized clustering methods to search optimal hypergraph structures and conducted convolution operation for representation learning. 
Also, some methods introduce the hypergraph structure into the recommendation task to mine beyond-pairwise relations among items~\cite{Wang@SDM2021, DHCN}. 

Considering that there are rich heterogeneous information and complex high-order relations in SBR, we combine the merits of both heterogeneous graph and hypergraph to build a novel heterogeneous hypergraph architecture. The customized heterogeneous hypergraph endows the proposed \baby with the capacity to extract price and interest preference of users. 

\subsection{Multi-task Learning}
Multi-task learning aims to handle multiple relevant tasks jointly via exploring mutual relations among these tasks~\cite{zhang@TKDE2021}. It treats all tasks equally and leverages the knowledge learned from each task to improve the model performance on different tasks. Recently, there a few works attempting to combine recommendation task with other auxiliary tasks to improve the accuracy for user modeling. In light of similar items with distinct IDs may reflect same user intent, Liu~\etal~\cite{Liu@WWW2020} view item text as additional signals to learn shared intent within similar items. Shalaby~\etal~\cite{Shalaby@RecSys2022} introduced the task of predicting item category to improve the model's predictive performance. In this work, considering that price and interest preference influence each other and collectively determine user choice, we draw support from multi-task learning to deduce these two preferences simultaneously and reveal user purchase intent.

\section{Motivation and Preliminaries}\label{sec:motivation}
\begin{figure}[t]
\centering
    \subfigure[{price distribution for users on different categories}]{
        \includegraphics[width=0.45\linewidth]{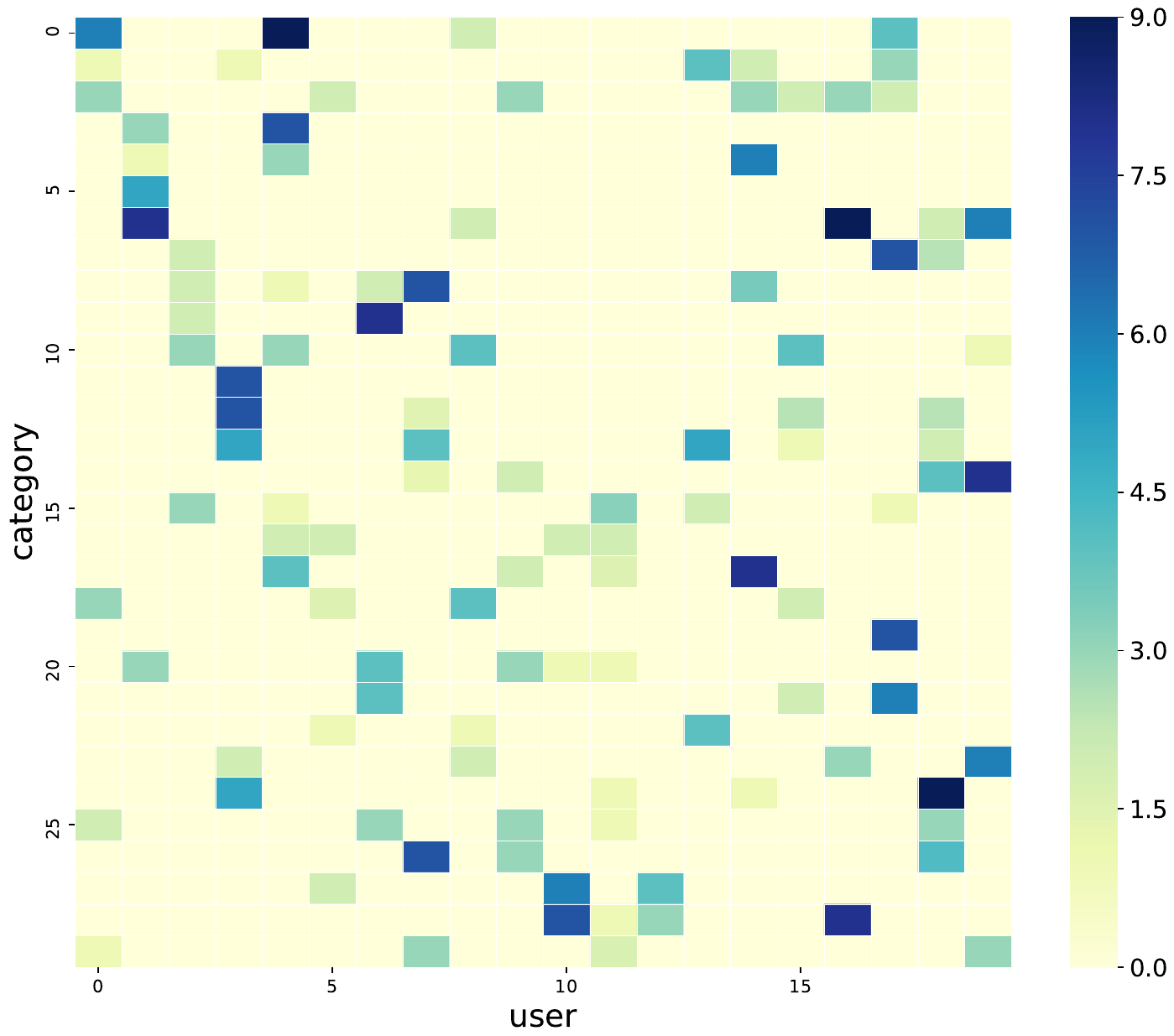}
    }
    \subfigure[{price distribution for users on different brands}]{
        \includegraphics[width=0.45\linewidth]{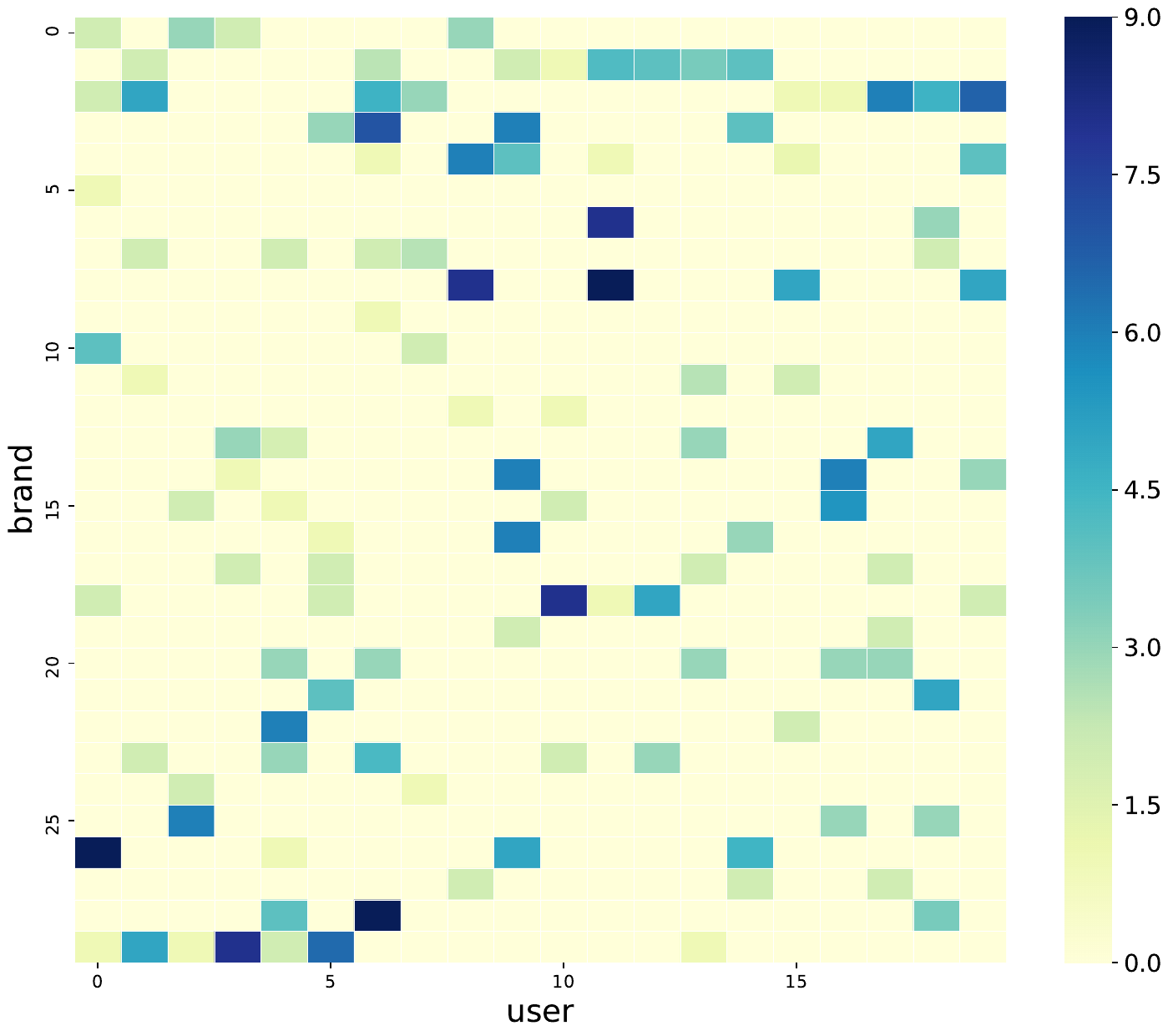}
    }
    \caption{Price heatmaps for users on different categories or brands. The depth of the color means how much money a user is will to spend on the corresponding categories/brands.}\label{priceHeatmap}
\end{figure}

\subsection{Motivation Study}\label{subsec:motivation}
In this part, we conduct data analysis on 'Grocery and Gourmet Food' dataset from Amazon to examine the relations between user price preference and categories/brands (ref. Section~\ref{sec:dataset} for details of datasets). In order to clearly display characteristics of data distribution, we randomly select 20 users who purchased at least 20 items and interacted with more than 4 categories/brands from the dataset. The average price levels on each category/brand for a user is used to indicate her price preference in the category/brand (ref. Section \ref{sec:pricedis} for price discretization). We present the price heatmap for users on different categories/brands at \fig~\ref{priceHeatmap}, where the darker the color is, the more money the user is willing to spend on corresponding categories/brands.

As shown in \fig ~\ref{priceHeatmap} (a) and (b), a user acceptable price level is extremely diverse among different categories/brands. That is, a user can be willing to buy expensive items in some categories/brands, while buying cheap items in others. It indicates that the item category/brand does influence the user price preference. As a result, it is rational to consider these two factors when modeling user price preference. Moreover, different users show different price preferences for a category/brand. As we can see from \fig ~\ref{priceHeatmap} (a), for a certain category, some users would like to pay high prices, while others would not. It suggests that there is divergence in user price preference, which motivates us to incorporate price into SBR for offering personalized services.

\subsection{Problem Statement}
This work aims to improve SBR by exploring user price and interest preference from various item features simultaneously. $\mathcal{I}$ is the item set, where $ n =|\mathcal{I}|$ is the total number of items users have purchased. In general, some features are available for an item $x_i$ $\in$ $\mathcal{I}$, such as item ID ($x_i^{id}$), item price($x_i^p$), item category($x_i^c$) and item brand($x_i^b$). In SBR, an anonymous user have purchased $m$ items during a certain period, generating a session $\mathcal{S}$ = [$x_1, x_2, ..., x_m$], where items are chronologically-ordered. Our goal is to predict the next item $x_{m+1}$ the user would like to buy based on $\mathcal{S}$. The main notations used in this work are detailed in Table~\ref{notation}.

\begin{table}[t]
\tabcolsep 0.01in 
\centering
\caption{Main Notations Used in This Article.}
\begin{tabular}{ll}
\toprule
\makecell[c]{Notation}      & \makecell[c]{Description} \cr
\midrule
$\mathcal{I}$, $n$/|$\mathcal{I}$|       & item set, the total number of items  \cr
$x_i$ & an item      \cr
$\mathcal{S} = [x_1, x_2, ..., x_m]$    & a session with $m$ items generated by an anonymous user     \cr
$min$, $max$       & min/max item price in a category     \cr
$\Phi(*)$ & cumulative distribution function of logistic distribution \cr
$round(*)$     & rounding operation \cr
$V$ and $E$    &  heterogeneous node set and hyperedge set   \cr
$V^{id}$, $V^{p}$, $V^{c}$ and $V^{b}$    &  node sets with type ID, price, category and brand   \cr
$p_j$, $c_k$ and $b_l$    &  a node with type price, category and brand    \cr
$\mathbf{v}_i^{id}$, $\mathbf{v}_j^p$, $ \mathbf{v}_k^c$ and $\mathbf{v}_l^b$    &    node embeddings with type ID, price, category and brand  \cr
$\mathbf{h}_i^{id}$, $\mathbf{h}_j^{p}$, $\mathbf{h}_k^{c}$ and $\mathbf{h}_l^{b}$ & updated embeddings for ID, price, category and brand nodes   \cr
$avg(*)$   &  average pooling operation   \cr
$f_{a}(*)$ and $f_{b}(*)$ & intra-type and inter-type convolution operation \cr
$\mathbf{u}_\mathsf{p}$ and $\mathbf{u}_\mathsf{I}$    &   the user price preference and interest preference  \cr
$\rho$ & the number of price levels \cr
$r$ & the repeating number of triple-level convolution \cr
$h$ & the number of self-attention heads
\\
                
\bottomrule
\end{tabular}
\label{notation}
\end{table}

\subsection{Price Discretization}\label{sec:pricedis}

\begin{figure}[t]
  \centering
  \includegraphics[width=0.6\linewidth]{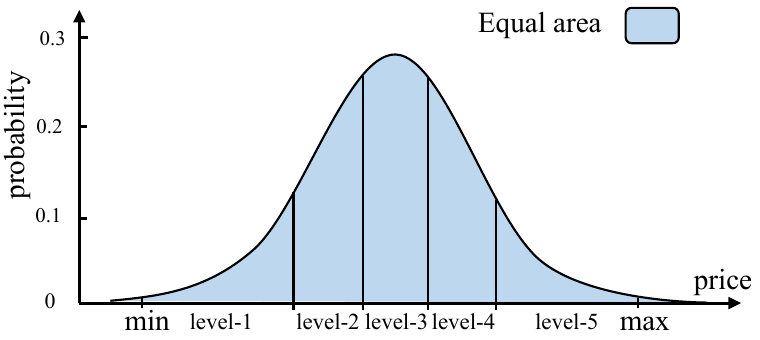}
  \caption{Logistic distribution. We discretize absolute price into different price levels by equally partitioning price probability distribution in a category. }\label{logisticDistri}
\end{figure}

As indicated in~\cite{Zheng@ICDE2020}, we can't estimate whether an item is expensive or cheap by its absolute price. Taking an example, \$300 is cheap for a computer, but it could be extremely expensive for a T-shirt. Apparently, to intuitively compare the item price from distinct categories, we need to divide absolute price into price levels based on item category. Asnat \etal~\cite{KBS2018price} found that the logistic distribution is more compatible with price distribution on a category than commonly used uniform distribution. The logistic probability density function is plotted in \fig \ref{logisticDistri}, where the curve is high in the middle and low on both sides. In the middle, there are more alternatives with similar price for users to select, leading to higher user sensitivity for price, which requires us to discretize price levels more finely. For both sides, the reverse applies. Moreover, neural models generally achieve good performance under the situation of balanced data. Thus, the size of each price level should be similar. In light of the aforementioned factors, as shown in \fig ~\ref{logisticDistri}, we divide price into $\rho$ levels (\eg $\rho=5$) by making the probability of every interval equal. More specifically, for an item whose absolute price is $x_p$ and $[min, max]$ is the min/max price among its category, its price level $p_i$ is set as follows,
\begin{align}
    p_i &= round(\frac{\Phi(x_p)-\Phi(min)}{\Phi(max)-\Phi(min)}\times\rho),
\end{align}
where $\Phi(x)$ denotes the cumulative distribution function for logistic distribution, and we can formally defined $\Phi(x)$ as follows,
\begin{align}
    \Phi(x) &= P(X \leq x) = \frac{1}{1 + e^{-\pi\frac{x-\mu}{\sqrt{3}\delta}} } ,
\end{align}
where $\mu$ is expected value and $\delta$ is standard deviation both of which are estimated by maximum likelihood in this work. 

\subsection{Heterogeneous Hypergraph Construction}
Let $\mathcal{G}$ = $(V, E)$ be the proposed heterogeneous hypergraph which consists of node set $V$ and hyperedge set $E$. A node $ {x}_i^{\tau} \in V$ has a type $\tau$ and nodes belonging to the same type $\tau$ constitute a homogeneous node set $V^{\tau}$. We consider four types of information in this paper, \ie item ID ($V^{id}$), item price ($V^{p}$), item category ($V^{c}$) and item brand ($V^{b}$), which are directly relevant to user price and interest preference. That is, $V = V^{id} \bigcup V^{p} \bigcup V^{c} \bigcup V^{b}$. A hyperedge $\epsilon \in E$ can contain any number of nodes with any types. Three types of hyperedges are designed to encode rich correlations among nodes: (1) a \textit{feature hyperedge} contains four features of the purchased item; (2) a \textit{price hyperedge} contains the price nodes appeared in a session. (3) a \textit{session hyperedge} contains the ID nodes within a session. Note that feature hyperedges are used to aggregate heterogeneous information in the heterogeneous hypergraph. We utilize the price/session hyperedges to mine user price/interest preference. If two nodes appear in any hyperedge simultaneously, they are defined as having adjacent relations. The leftmost part of \fig ~\ref{model} shows a toy example that illustrates how we construct the heterogeneous hypergraph.

\section{The Proposed Approach}\label{sec:method}

\subsection{Overview of the \baby}
The schematic illustration of the proposed \baby is presented in \fig~\ref{model}. On the basis of the customized heterogeneous hypergraph, a novel triple-level convolution is designed to aggregate heterogeneous information from three levels, \ie co-occurrence level, intra-type level and inter-type level, for node embedding learning. We then apply attention layers with temporal information to obtain price and interest preference of users. After that, a bi-preference learning schema is devised to reveal user purchase intent via predicting price and interest preference simultaneously. In following parts, we will describe each component in detail.

\subsection{Triple-level Convolution}\label{sec:triple}
\begin{figure}[t]
  \centering
  \includegraphics[width=0.95\linewidth]{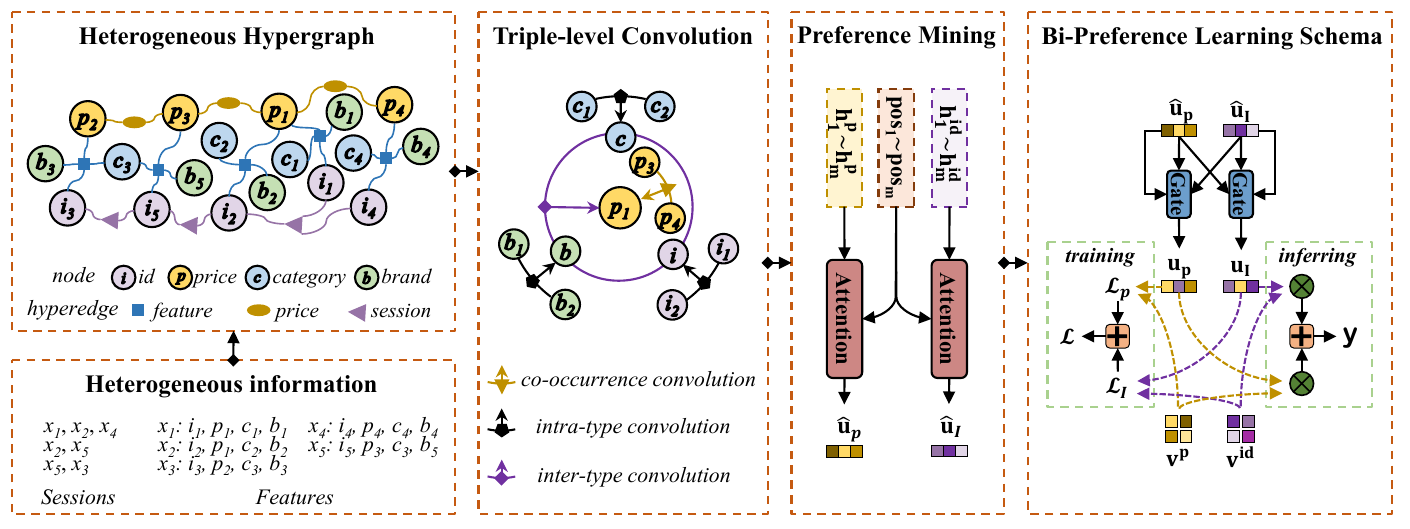}
  \caption{Schematic illustration of the proposed \baby. Based on the customized heterogeneous hypergraph, the triple-level convolution is proposed to learn node embeddings via mining co-occurrence, intra-type and inter-type relations among various nodes. Two attention layers are then utilized to mine user price and interest preference. Finally, the bi-preference learning schema is devised to jointly predict user price and interest preference by exploring multual relations between these two preferences. (\textit{Best viewed in color})}\label{model}
\end{figure}

For a target node in the heterogeneous hypergraph, there are multiple relations we should consider to build its embedding. On the one hand, as demonstrated in most recommendation algorithms~\cite{BPR-MF, Koren@KDD2008, GRU4Rec, SR-GNN}, items purchased together possess similar characteristics, \ie there are \textit{co-occurrence} relations in these items. In our settings, for a target node (\eg $p_1$ in \fig~\ref{model}), the nodes with same type appeared in a session ({$ p_3, p_4 $}) have similar semantics and can help model understand its meaning. On the other hand, for the target node ($p_1$), the adjacent nodes with other types (\eg $\{c_1, c_2\} \to p_1$, $\{b_1, b_2\} \to p_1$ and $\{i_1, i_2\} \to p_1 $) can also provide useful information~\cite{Hu@EMNLP2019}. More concretely, the nodes with a certain type (\eg category) contain homogeneous information, where different nodes ($\{c_1, c_2\}$) may have different importance to the target node, \ie \textit{intra-type} relations. Also, different types furnish heterogeneous information to the target node (\eg category $\to$ price, brand $\to$ price or ID $\to$ price), \ie \textit{inter-type} relations. Therefore, we propose a novel triple-level convolution to learn node embeddings via collectively investigating these three types of relations on the heterogeneous hypergraph, \ie \textit{co-occurrence convolution}, \textit{intra-type convolution} and \textit{inter-type convolution}.

\paratitle{Co-occurrence convolution.}
The goal of the co-occurrence convolution is to enhance the node embeddings via exploring co-occurrence relations. Let $\mathbf{v}_i^{t} \in \mathbb{R}^d$ be the embedding of the target node $x_i^t$ with type $t$. The $N_i^t$ is the set of nodes whose type is $t$ and having co-occurrence relations with $x_i^t$. For the simplicity and efficiency, the co-occurrence convolution applies the average pooling operation to extract co-occurrence information as follows, 
\begin{align}
    \mathbf{c}_{i} &= avg(N_{i}^{t}) .
\end{align}

\paratitle{Intra-type convolution.}
The intra-type convolution attempts to distinguish the significance of nodes within a certain type, so as to extract relevant information of this type about the target node. For a target node $\mathbf{v}_i^{t} $, the $N_{i}^{\tau}$ is the set of its adjacent nodes with the type $\tau$. The intra-type convolution prioritizes the nodes within $ N_{i}^{\tau}$ to learn the specific type embedding about $\mathbf{v}_i^{t} $ for the type $\tau$ as follows,
\begin{align}
    \mathbf{e}_{i}^{\tau} &= \sum_{k} \alpha_k \mathbf{v}_k^{\tau},\\
    \alpha_k &= \frac{exp(\mathbf{v}_i^{t \top}  \mathbf{W} \mathbf{v}_k^{\tau})}{\sum_{\mathbf{v}_j^{\tau} \in N_{i}^{\tau}}\exp(\mathbf{v}_i^{t \top}  \mathbf{W}  \mathbf{v}_j^{\tau})},
\end{align}
where $\mathbf{W} \in \mathbb{R}^{d \times d}$ is the learnable parameter used to evaluate the similarity between the target node $\mathbf{v}_i^{t}$ and $\mathbf{v}_k^{\tau} \in N_{i}^{\tau}$. The $\mathbf{e}_{i}^{\tau}  \in \mathbb{R}^{d} $ represents the information which is needed to propagate from type $\tau$ to $\mathbf{v}_i^{t}$. Considering that the nodes belonging to the same type possess homogeneous information, the intra-type convolution aggregates the information via linearly weighted addition as in Equation(4). For the convenience of the following descriptions, we simplify the intra-type convolution formulated by Equation(4-5) as follows,
\begin{align}
    \mathbf{e}_{i}^{\tau} &= f_{a}(N_{i}^{\tau}).
\end{align}

\paratitle{Inter-type convolution.}
The inter-type convolution aims to integrate relevant heterogeneous information from different types into the target node. Considering that different types offer distinct semantics to the target node, the inter-type convolution fuses the type embeddings via gating operations as follows,
\begin{align}
    \mathbf{h}_i &= \mathbf{v}_i^t + \mathbf{g}_{1} \odot \mathbf{e}_{i}^{\tau 1} + \mathbf{g}_{2} \odot \mathbf{e}_{i}^{\tau 2} + \mathbf{g}_{3} \odot \mathbf{e}_{i}^{\tau 3}, \\
    \mathbf{g}_{1} &= tanh (\mathbf{e}_{i} + \mathbf{W}_{1}\mathbf{e}_{i}^{\tau 1}), \\
    \mathbf{g}_{2} &= tanh (\mathbf{e}_{i} + \mathbf{W}_{2}\mathbf{e}_{i}^{\tau 2}), \\
    \mathbf{g}_{3} &= tanh (\mathbf{e}_{i} + \mathbf{W}_{3}\mathbf{e}_{i}^{\tau 3}), \\
    \mathbf{e}_{i} &= \mathbf{W}_{a}[\mathbf{v}_i^t;\mathbf{e}_{i}^{\tau 1};\mathbf{e}_{i}^{\tau 2};\mathbf{e}_{i}^{\tau 3}],
\end{align}
where $\mathbf{W}_a \in \mathbb{R}^{d \times 4d}$, $\mathbf{W}_{1}$, $\mathbf{W}_{2}$ and $\mathbf{W}_{3} \in \mathbb{R}^{d \times d} $ are learnable parameters, $\odot$ is element-wise product, $ tanh(*) $ is activate function and [;] denotes concatenation. The $\mathbf{W}_a$ merges various heterogeneous information into $\mathbf{e}_{i}$ which is used to guide the information fusing next. We then utilize gating mechanism to obtain the embedding of target node $\mathbf{h}_i$. Noting that the semantics of $\mathbf{h}_i$ are enriched by rich heterogeneous information, which enables \baby to perceive various preferences of users. We can simplify the inter-type convolution defined by Equation(7-11) as, 
\begin{align}
    \mathbf{h}_i &= f_{b}(\mathbf{v}_i^t, \mathbf{e}_{i}^{\tau 1}, \mathbf{e}_{i}^{\tau 2}, \mathbf{e}_{i}^{\tau 3}).
\end{align}

As detailed above, the triple-level convolution conducts information aggregating on the heterogeneous hypergraph by exploring co-occurrence relations, intra-type relations and inter-type relations existing within heterogeneous nodes. Formally, the triple-level convolution updates node embeddings with four types, \ie $\mathbf{v}_i^{id} \in V^{id}$, $\mathbf{v}_j^p \in V^{p}$, $ \mathbf{v}_k^c \in V^{c}$ and $\mathbf{v}_l^b \in V^{b}$, as follows,
\begin{align}
\mathbf{h}_i^{id} &= f_{b}(\mathbf{v}_i^{id}, f_{a}(N_{i}^p), f_{a}(N_{i}^c), f_{a}(N_{i}^b)) + \mathbf{c}_i, \\
\mathbf{h}_j^{p} &= f_{b}(\mathbf{v}_j^{p}, f_{a}(N_{j}^{id}), f_{a}(N_{j}^c), f_{a}(N_{i}^b)) +  \mathbf{c}_j, \\
\mathbf{h}_k^{c} &= f_{b}(\mathbf{v}_k^{c}, f_{a}(N_{k}^{id}), f_{a}(N_{k}^p),  f_{a}(N_{k}^b)) + \mathbf{c}_k, \\
\mathbf{h}_l^{b} &= f_{b}(\mathbf{v}_l^{b}, f_{a}(N_{l}^{id}), f_{a}(N_{l}^p), f_{a}(N_{l}^c)) + \mathbf{c}_l,
\end{align}
where $\mathbf{h}_i^{id}$, $\mathbf{h}_j^{p}$, $\mathbf{h}_k^{c}$ and $\mathbf{h}_l^{b} \in \mathbb{R}^{d}$ are updated embeddings with type of ID, price, category and brand respectively. With repeating the triple-level convolution $r$ times, we can handle heterogeneous information to model user price and interest preference. 

\subsection{Preference Mining}\label{sec:preference}
After obtaining node embeddings, we further mine a user price and interest preference exposed in a session to predict her future behaviors.

\paratitle{Price preference mining.}
Intuitively, the price preference of a user is hidden in the item price within the session. Therefore, we mine user price preference based on price hyperedge, \ie price sequence $ [\mathbf{h}_1^{p},  \mathbf{h}_2^{p},  \mathbf{h}_3^{p}, ...,  \mathbf{h}_m^{p}]$. Given that the user price preference may change alone with time, we first introduce temporal information into the price embedding to encode the dynamics. Specifically, the reversed position embeddings \cite{GCE-GNN,DHCN} is used to represent the temporal information, \ie $\mathbf{pos}_i \in \mathbb{R}^{d}$. We then integrate the price embedding with position embedding as follows,
\begin{align}
     \mathbf{h}_i^*  = tanh(\mathbf{W}_p[\mathbf{h}_i^{p};\mathbf{pos}_i] + \mathbf{b}_p),
\end{align}
where $\mathbf{W}_p \in \mathbb{R}^{d \times 2d}$ and $\mathbf{b}_p \in \mathbb{R}^d$ are learnable parameters. $\mathbf{h}_i^* \in \mathbb{R}^d$ is the modified embedding of price node whose semantics are enriched by temporal information. Afterwards, considering that the self-attention mechanism is good at modeling transition relations within a sequence for semantics capturing~\cite{SASRec, GC-SAN, BERT4Rec}, we employ multi-head self-attention to extract user price preference as follows,
\begin{align}
    \mathbf{E}_{p} &= [\mathbf{h}_1^*;\mathbf{h}_2^*; ...; \mathbf{h}_m^*], \\
    head_i &= Attention(\mathbf{W}_i^Q\mathbf{E}_p, \mathbf{W}_i^K\mathbf{E}_p, \mathbf{W}_i^V\mathbf{E}_p), \\
    \mathbf{S}_p &= [head_1; head_2; ...; head_h],
\end{align}
where $\mathbf{W}_i^Q, \mathbf{W}_i^K$ and $\mathbf{W}_i^V \in \mathbb{R}^{\frac{d}{h} \times d} $ are learnable parameters which are used to convert the input to query, key and value space respectively, and $h$ is the number of self-attention heads. After processing the price sequence through multi-head self-attention, we use the hidden state $\mathbf{S}_p$ to represent a user price preference, \ie $\mathbf{\hat{u}}_\mathsf{p} \in \mathbb{R}^d$, as follows,
\begin{align}
     \mathbf{\hat{u}}_\mathsf{p} &= \mathbf{S}_p^{(m)}.
\end{align}

\paratitle{Interest preference mining.}
We rely on the session hyperedge, \ie item ID sequence $ [ \mathbf{h}_1^{id},  \mathbf{h}_2^{id}, \mathbf{h}_3^{id}, ...  \mathbf{h}_m^{id}]$, to mine user interest preference, since that it is the items a user has bought that express her interest~\cite{NARM, LESSR, DHCN}. The user interest preference always dynamically changes according to time~\cite{Wang@CS2022,DHCN,DIDN}, so we also use the position embeddings ($\mathbf{pos}_i$) to enhance item ID embeddings as follows,
\begin{align}
     \mathbf{v}_i^*  = tanh(\mathbf{W}_f[\mathbf{h}_i^{id};\mathbf{pos}_i] + \mathbf{b}_f),
\end{align}
where $\mathbf{W}_f \in \mathbb{R}^{d \times 2d}$ and $\mathbf{b}_f \in \mathbb{R}^d$ are learnable parameters. As defined by Equation(22), the $i$-th item in a session is represented by  $\mathbf{v}_i^* \in \mathbb{R}^d$. Following the common method to formulate user interest preference~\cite{STAMP,SR-GNN}, we define the interest preference of a user as follows,
\begin{align}
    \mathbf{\hat{u}}_\mathsf{I} &= \sum^{m}_{i=1} \beta_i \mathbf{h}_i^{id}, \\
    \beta_i &= \mathbf{z}\sigma(\mathbf{A}_1\mathbf{v}_i^{*}+\mathbf{A}_2\mathbf{\bar{v}}^{*}+\mathbf{b}),
\end{align}
where $\mathbf{A}_1$, $ \mathbf{A}_2  \in \mathbb{R}^{d \times d}$ and $\mathbf{b \in \mathbb{R}^d}$ are learnable parameters, $\mathbf{z}^T \in \mathbb{R}^d$ is the attention matrix used to determine the importance of an item within the session. The $\mathbf{\bar{v}}^{*} = \frac{1}{m}\sum_{i=1}^m \mathbf{v}_i^*$ is the average embeddings of all item ID within the session, which is utilized to evaluate the contribution of various items to user interest preference. Consequently, the user interest preference is represented as $\mathbf{\hat{u}}_\mathsf{I} \in \mathbb{R}^d$.

\subsection{Bi-Preference Learning Schema}
As discussed before, both price preference and interest preference are indispensable in SBR. There exists rich mutual relations between these two preferences, leading to their joint determination for user purchase decisions. The multi-task learning mechanism can enhance the model capacity by leveraging the mutual information among multiple relevant tasks~\cite{zhang@TKDE2021}. Inspired by this, we develop a bi-preference learning schema under the multi-task learning architecture to collectively predict price and interest preference. We first enrich the semantics of price preference $\mathbf{\hat{u}}_\mathsf{p}$ and interest preference $\mathbf{\hat{u}}_\mathsf{I}$ as follows,
\begin{align}
    \mathbf{u}_\mathsf{p} &= \mathbf{r}_\mathsf{p} * \mathbf{\hat{u}}_\mathsf{p} + (1-\mathbf{r}_\mathsf{p}) * \mathbf{\hat{u}}_\mathsf{I}, \\
    \mathbf{u}_\mathsf{I} &= \mathbf{r}_\mathsf{I} * \mathbf{\hat{u}}_\mathsf{I} + (1-\mathbf{r}_\mathsf{I}) * \mathbf{\hat{u}}_\mathsf{p}, \\
    \mathbf{r}_\mathsf{p} &= \sigma(\mathbf{W}_1^{p}(\mathbf{\hat{u}}_\mathsf{p}) + \mathbf{W}_2^{p}(\mathbf{m})),\\
    \mathbf{r}_\mathsf{I} &= \sigma(\mathbf{W}_1^{I}(\mathbf{\hat{u}}_\mathsf{I}) + \mathbf{W}_2^{I}(\mathbf{m})),\\
    \mathbf{m} &= tanh(\mathbf{W}_1^{pi}(\mathbf{\hat{u}}_\mathsf{p}) + \mathbf{W}_2^{pi}(\mathbf{\hat{u}}_\mathsf{I}) + \mathbf{b}_{pi}),
\end{align}
where $\mathbf{W}_*^{pi}$, $\mathbf{W}_*^{P}$, $\mathbf{W}_*^{I} \in \mathbb{R}^{d \times d}$,  $\mathbf{b}_{pi} \in \mathbb{R}^{d}$ are learnable parameters. The $\mathbf{m} \in \mathbb{R}^d$ mergers the price and interest preferences to guide the fusion of these two kinds of information next. Referring to the gating mechanism~\cite{GRU4Rec}, we set the 'remember gate', \ie $\mathbf{r}_\mathsf{p}$/$\mathbf{r}_\mathsf{I}$ $ \in \mathbb{R}^d$, to determine the degree of the price/interest information needed to retain when merging these two preferences. After that, we further enhance the representations of these two preferences ($\mathbf{u}_\mathsf{p}$ and $\mathbf{u}_\mathsf{I}$) by exploring mutual relations between them with the help of multi-task learning where user price preference and interest preference are deduced simultaneously. 

\paratitle{Interest preference prediction.} As common practices in previous works~\cite{GRU4Rec,NARM,BERT4Rec,SR-GNN,LESSR,Xia@CIKM2021,Guo@WSDM2022}, we can formulate the task of user interest preference prediction based on user interest embedding $\mathbf{u}_\mathsf{I}$ and item ID embedding $\mathbf{v}_i^{id}$.
Formally, we predict user interest preference based on $\mathbf{u}_\mathsf{I}$ and $\mathbf{v}_i^{id}$ as follows,
\begin{align}
    y^I_i &= softmax(\mathbf{u}_\mathsf{I}^{\top}  \mathbf{v}_i^{id}), \\
    \mathcal{L}_{I} &= - \sum^n_{j=1} p^I_j \log (y^I_j),
\end{align}
where $y^I_i$ is predicted probability of item $x_i$ to be liked by the user, $p^I$ is ground truth distribution of item ID, and $\mathcal{L}_{I}$ is the loss of interest preference prediction.

\paratitle{Price preference prediction.}  Both price and interest preferences play an important role in determining user choice. Thus, different from existing works~\cite{NARM,SR-GNN,LESSR,Guo@WSDM2022} that only take user interest preference into consideration, \baby emphasizes the significance of user price preference. To be specific, besides predicting user interest, \baby also introduces the item price as extra signals to formulate the task of price preference prediction. Given the learned representation of user price preference $\mathbf{u}_\mathsf{p}$ and item price $\mathbf{v}_i^{p}$, we can utilize the price level of next purchased item to formulate the price preference prediction task as follows,
\begin{align}
    y^p_i &= softmax(\mathbf{u}_\mathsf{p}^{\top}  \mathbf{v}_i^p), \\
    \mathcal{L}_{p} &= - \sum^n_{j=1} p^p_j \log (y^p_j),
\end{align}
where $y^p_i$ is predicted probability of item price level, $p^p$ is ground truth price level distribution, and $\mathcal{L}_{p}$ is the loss of price preference prediction.

\subsection{Model Training and Inferring}

\paratitle{Training phase.} At training phase, we explore rich mutual relations between price and interest preference by deducing these two preferences simultaneously. Concretely, we jointly consider the tasks of price preference prediction and interest preference prediction to reveal user purchase intent as follows,
\begin{align}
    \mathcal{L} &=  \mathcal{L}_{p} + \mathcal{L}_{I}.
\end{align}

Note that, different from common paradigm of multi-task learning that balances the importance of different tasks with a constant, we directly add losses of these two tasks as final loss. It implies that we view the price and interest preference to be equally crucial in affecting user behaviors.

\paratitle{Inferring phase.} As discussed in former sections, a user purchase behaviors is determined by both her price preference and interest preference on an item. Thus, at inferring phase, we should consider these preferences collectively. Based on the learned user preferences ($\mathbf{u}_\mathsf{p}$ and $\mathbf{u}_\mathsf{I}$) and item features ($\mathbf{v}_i^{p}$ and $\mathbf{v}_i^{id}$), the purchased probability of an item can be calculated as follows,
\begin{align}
    y_i &= softmax(\mathbf{u}_\mathsf{p}^{\top}  \mathbf{v}_i^p + \mathbf{u}_\mathsf{I}^{\top}  \mathbf{v}_i^{id}),
\end{align}
where the $y_i$ represents the probability the user buys item $x_i$.

\section{Experimental Settings}\label{sec:experimentalsetting}
\subsection{Research Questions}
In order to examine \baby's performance in SBR, we conduct a number of experiments on three public real-world datasets. Specifically, we aim to answer following research questions:
\begin{itemize}
    \item \textbf{RQ1} Does the proposed \baby achieve state-of-the-art performance compared with competitive SBR baselines? (ref. Section~\ref{sec:overall})
    
    \item \textbf{RQ2} Does each design choice contribute positively to the performance of the proposed \baby? (ref. Section~\ref{sec:price}-~\ref{sec:bipre})

    \item \textbf{RQ3} How does the proposed \baby perform under different price levels? (ref. Section~\ref{sec:priceLevel})

    \item \textbf{RQ4} What's the performance of SBR under various session length?  (ref. Section~\ref{sec:sessionlength})

    \item \textbf{RQ5} How does the key hyper-parameters influence \baby? (ref. Section~\ref{sec:hyper})
    
    \item \textbf{RQ6} How does various information influence \baby's complexity? (ref. Section~\ref{sec:complexity})

\end{itemize}

\subsection{Datasets and Preprocessing}\label{sec:dataset}

\begin{table}[t]
\tabcolsep 0.1in 
\centering
\caption{Statistics of Datasets.}
\renewcommand{\arraystretch}{1.1}
\begin{tabular}{ccccc}
\toprule
Datasets      &Cosmetics & Grocery & Toys \cr
\midrule
\#item        & 13,026  & 6,230     & 18,979     \cr
\#price level & 10      & 5         & 5        \cr
\#category    & 226     & 550       & 430      \cr
\#brand       & 148     & 1,306     & 1,429      \cr
\#interaction & 654,947 & 365,415   & 714,770    \cr
\#session     & 109,251 & 153,383   & 287,733     \cr
avg.length    & 5.99    & 2.38      & 2.48      \cr
\bottomrule
\end{tabular}
\label{statistics}
\end{table}

We use following three public real-world datasets in this work:
\begin{itemize}
    \item \textbf{Cosmetics} \footnote{\url{https://www.kaggle.com/mkechinov/ecommerce-events-history-in-cosmetics-shop}}, as a public dataset in kaggle platform, traces user actions in an online cosmetics shop. The records of a month (October 2019) are used in this work, where we choose the interactions with type 'purchase' and 'add\_to\_cart' to formulate user purchase behaviors.
    
    \item Amazon \footnote{\url{http://jmcauley.ucsd.edu/data/amazon/}} records rich characteristics of users and items from the popular e-commerce website Amazon \cite{Amazon}. We select two representative sub-datasets 'Grocery and Gourmet Food' (\textbf{Grocery}) and 'Toys and Games' (\textbf{Toys}) in this work. To simulate the scenario of SBR, the user behaviors happened within one day are formulated as a session. 
    
\end{itemize}

As the common settings in SBR~\cite{NARM,SR-GNN,LESSR,DHCN,Xia@CIKM2021}, we delete the sessions which only contain one item in all datasets. To guarantee data quality~\cite{Zheng@ICDE2020}, we apply 10-core settings to filter out data, where each item must appear at least 10 times and each price level/category/brand must contain at least 10 items. That is, \baby jointly explores item ID, price, category and brand to reveal user intents. We define the last item in a session as the ground truth and use the remaining actions to model user behaviors. The earliest 70\% sessions formulate the training set. And the next 20\% of sessions are used as validation set to determine the hyper-parameters of \baby and baselines. The remaining 10\% are used as test set to evaluate the model performance. Table~\ref{statistics} presents the detailed statistics of all datasets. 

\subsection{Evaluation Metrics}
In order to evaluate the performance of \baby and all baselines, we employ two widely used metrics Prec@k and MRR@k:
\begin{itemize}
\item Prec@k: Precision evaluates whether the ground truth item is in the recommendation list. More specifically, Prec@k measures the proportion of cases where the ground-truth item is within the top-$k$ recommendation list.

\item MRR@k: Mean Reciprocal Rank considers the position of the ground truth item in the recommendation list. Formally, MRR@k is the mean of the reciprocal rankings of ground truth item in the list. If the rank is more than k, the reciprocal rank is manually set to zero.
\end{itemize}

Note that, Prec@k does not take the item's actual rank into account as long as it is among the recommendation list. In contrast, the item rank is taken into consideration by MRR@k, which is significant when the recommended order matters.
In addition, for both Prec@k and MRR@k, larger numbers indicate better model performance. According to conventional settings~\cite{NARM,BERT4Rec, LESSR,Xia@CIKM2021}, the k is set as 10 and 20 in this work.

\subsection{Comparison Methods}
Following competitive methods are adopted as baselines to examine the performance of the proposed \baby:
\begin{itemize}
    \item \textbf{S-POP} recommends the most popular items within the current session.
    
    \item \textbf{SKNN} evaluates the similarity between current session and past sessions to determine the recommended items. 
    \item \textbf{GRU4Rec}~\cite{GRU4Rec} captures sequential relations within the session via GRU.
    \item \textbf{NARM}~\cite{NARM} improves GRU4Rec by introducing attention mechanism to mine user main intention. 
    \item \textbf{BERT4Rec}~\cite{BERT4Rec} uses a bidirectional self-attention to model user sequential behaviors. 
    \item \textbf{SR-GNN}~\cite{SR-GNN} views each session as a graph and applies graph neural networks on the graph to model pair-wise relations between items.
    \item \textbf{SR-GNN+} modifies SR-GNN by adding the embeddings of item price, category and brand as inputs to examine the effect of side information on SBR.
    \item \textbf{LESSR}~\cite{LESSR} tackles lossy session encoding and ineffective long range dependency of GNN for SBR.
    \item \textbf{$S^2$-DHCN}~\cite{DHCN} introduces hypergraph architecture into SBR and further improve performance by contrastive learning.
    \item \textbf{COTREC}~\cite{Xia@CIKM2021} explores internal and external connectivity of sessions with self-supervised learning and co-training.
    \item \textbf{MSGIFSR}~\cite{Guo@WSDM2022} captures user fine-grained preferences evolution by modeling multi-granularity consecutive user intent.
    \item \textbf{MGS}~\cite{MGS} exploits item attribute information (\eg categories and brands) for accurate preferences learning.
    \item \textbf{DGNN}~\cite{Li@WSDM2023} mines explicit and implicit item relations with GNN to improve SBR.
    \item \textbf{CoHHN}~\cite{CoHHN} is a preliminary version of this paper, which extracts user price preference from item price and category.
    
\end{itemize}

\subsection{Implementation Details}
For fair comparison, we fix embedding size at 128 for all neural methods. We determine other hyper-parameters of \baby and baselines by grid search based on their performance at Prec@20 on validation set\footnote{{The settings of neural baseline methods are as follows: NARM (1 GRU layer with 128 hidden units), BERT4Rec (4 heads and 2 layers), LESSR (4 layers), $S^2$-DHCN (3 layers), COTREC (3 layers), MSGIFSR (2 layers and 3 intent granularity levels), MGS (3 samples in mirror graph), DGNN (2 layers).}}. 
As to the main hyper-parameters of \baby, we investigate the number of price levels $\rho$ in $\{2, 5, 10, 50, 100, 200\}$, the repeating number of triple-level convolution $r$ in $\{1, 2, 3, 4\}$ and the number of self-attention heads $h$ in $\{1, 4, 8, 16, 32\}$. Besides, we set the mini-batch size at $100$ and optimize the model via Adam where the initial learning rate is $0.001$. To simulate the actual scenarios of SBR as in~\cite{NARM,SR-GNN,LESSR}, the length of session is cut off at $19$. We run every model five times with random initial seeds and use the average performance on text sets as final results for all approaches. The source codes are available online.\footnote{{https://github.com/Zhang-xiaokun/BiPNet}}

\section{Experiments}\label{sec:experiments}

\subsection{Overall Performance (RQ1)}\label{sec:overall} 

\begin{table}[t]
\small
\tabcolsep 0.05in 
    \caption{Overall Performance on Three Datasets.}
    \begin{tabular}{c cc cc cc}  
    \toprule  
    \multirow{2}*{Method}& 
    \multicolumn{2}{c}{Cosmetics}&\multicolumn{2}{c}{Grocery}&\multicolumn{2}{c}{Toys}\cr  
    \cmidrule(lr){2-3} \cmidrule(lr){4-5} \cmidrule(lr){6-7}
    &Prec@10&MRR@10 &Prec@10&MRR@10 &Prec@10&MRR@10\cr  
    \midrule  
    S-POP           &38.05&28.09	&38.19&33.31	&25.09&24.83\cr  
    SKNN            &42.52&31.22    &67.87&47.42    &45.49&32.16\cr  
    GRU4Rec         &21.29 $\pm$ 0.18 &15.61 $\pm$ 0.13    
    &59.38 $\pm$ 0.19 &53.78 $\pm$ 0.14    
    &35.99 $\pm$ 0.13  &28.44 $\pm$ 0.10 \cr  
    NARM            &43.99 $\pm$ 0.09 &34.18 $\pm$ 0.11   
    &69.68 $\pm$ 0.13  &62.54 $\pm$ 0.08    
    &45.92 $\pm$ 0.11 &37.43 $\pm$ 0.09 \cr
    BERT4Rec        &40.95 $\pm$  0.08 &24.43 $\pm$  0.07   
    &68.60 $\pm$ 0.18 &61.29 $\pm$ 0.14 
    &43.76 $\pm$ 0.17 &35.40 $\pm$ 0.14 \cr  
	SR-GNN          &45.87 $\pm$  0.10 &{34.52} $\pm$  0.05   
 &69.89 $\pm$ 0.14 &\underline{62.78} $\pm$ 0.11    
 &46.76 $\pm$ 0.15  &37.76 $\pm$ 0.09 \cr
     SR-GNN+          &46.03 $\pm$  0.12 &{34.49} $\pm$  0.11   
     &69.97 $\pm$ 0.12 &62.67 $\pm$ 0.15    
     &46.93 $\pm$ 0.11  &\underline{37.81} $\pm$ 0.13 \cr
	LESSR           &41.98 $\pm$  0.09 &26.07 $\pm$  0.11   
 &68.22 $\pm$ 0.12  &60.99 $\pm$ 0.16     
 &43.80 $\pm$ 0.09  &35.45 $\pm$ 0.11 \cr
	$S^2$-DHCN      &42.87 $\pm$  0.08 &33.19 $\pm$  0.12    
 &61.41 $\pm$ 0.13  &50.58 $\pm$ 0.11     
 &38.67 $\pm$ 0.12  &30.33 $\pm$ 0.08 \cr
	COTREC          &46.30 $\pm$  0.06 &34.72 $\pm$  0.08    
 &\underline{70.03} $\pm$ 0.08 &{51.29} $\pm$ 0.16   
 &\underline{47.49} $\pm$ 0.10 &33.59 $\pm$ 0.13 \cr
    MSGIFSR &43.70  $\pm$  0.11 &26.51 $\pm$  0.06	
    &69.38 $\pm$ 0.13  &61.64 $\pm$ 0.09 	
    &45.80 $\pm$ 0.16 &36.44 $\pm$ 0.09 \cr
    MGS & \underline{47.01} $\pm$ 0.15  & \underline{34.84} $\pm$ 0.09	
    &69.92 $\pm$ 0.12 &62.28 $\pm$ 0.18	
    &47.00 $\pm$ 0.08 &37.65 $\pm$ 0.11 \cr
    DGNN & 44.47 $\pm$ 0.18  & 33.86 $\pm$ 0.12	
    &69.55 $\pm$ 0.15 &62.63 $\pm$ 0.17	
    &45.30 $\pm$ 0.13 &37.59 $\pm$ 0.14 \cr
    \midrule  
    CoHHN           &47.95 $\pm$ 0.12 &35.02 $\pm$ 0.08 
    &70.54 $\pm$ 0.11 &63.16 $\pm$ 0.10	
    &47.89 $\pm$ 0.14 &38.54 $\pm$ 0.08 \cr
	{\bf \baby}     &{ $ \bf 48.64^*$}$\pm$ 0.14 &{$\bf 35.48^*$}$\pm$ 0.11
	                &{ $ \bf 71.12^*$}$\pm$ 0.10 &{$\bf 63.51^*$}$\pm$ 0.08
	                &{ $ \bf 48.48^*$}$\pm$ 0.09 &{$\bf 38.71^*$}$\pm$ 0.07 \cr
	\bottomrule
    \end{tabular}

    ~\\\

    \begin{tabular}{c cc cc cc}  
    \toprule  
    \multirow{2}*{Method}& 
    \multicolumn{2}{c}{Cosmetics}&\multicolumn{2}{c}{Grocery}&\multicolumn{2}{c}{Toys}\cr  
    \cmidrule(lr){2-3} \cmidrule(lr){4-5} \cmidrule(lr){6-7}
    &Prec@20&MRR@20 &Prec@20&MRR@20 &Prec@20&MRR@20\cr  
    \midrule  
    S-POP           &40.75&30.22	&41.30&33.52	&25.80&24.88\cr  
    SKNN            &49.83&32.36    &70.07&47.55    &48.63&32.38\cr  
    GRU4Rec         &23.88 $\pm$ 0.16 &15.79 $\pm$ 0.17    
    &62.53 $\pm$ 0.15 &54.03 $\pm$ 0.14    
    &38.24 $\pm$ 0.18 &29.01 $\pm$ 0.13 \cr  
    NARM            &48.52 $\pm$ 0.17 &34.50 $\pm$ 0.09 
    &71.54 $\pm$ 0.12 &62.81 $\pm$ 0.10   
    &48.42 $\pm$ 0.13 &37.81 $\pm$ 0.08\cr
    BERT4Rec        &49.40 $\pm$ 0.07 &25.01 $\pm$ 0.06 
    &71.01 $\pm$ 0.16 &61.45 $\pm$ 0.14
    &47.44 $\pm$ 0.14 &35.66 $\pm$ 0.10 \cr  
	SR-GNN          &50.75 $\pm$ 0.16 &34.78 $\pm$ 0.13   
 &71.68 $\pm$ 0.14 &63.03 $\pm$ 0.11   
 &49.48 $\pm$ 0.11 &38.05 $\pm$ 0.08 \cr
    SR-GNN+          &50.98 $\pm$ 0.12 &34.90 $\pm$ 0.17   
 &71.93 $\pm$ 0.16 &\underline{63.11} $\pm$ 0.13   
 &49.71 $\pm$ 0.09 &37.92 $\pm$ 0.11 \cr
	LESSR           &50.18 $\pm$ 0.07 &26.64 $\pm$ 0.11  
 &70.24 $\pm$ 0.09 &61.14 $\pm$ 0.08   
 &47.10 $\pm$ 0.12 &35.67 $\pm$ 0.07\cr
	$S^2$-DHCN      &49.79 $\pm$ 0.15 &33.33 $\pm$ 0.09    
 &63.28 $\pm$ 0.17 &50.87 $\pm$ 0.12    
 &41.18 $\pm$ 0.09 &30.61 $\pm$ 0.10 \cr
	COTREC          &51.41 $\pm$ 0.13 &34.93 $\pm$ 0.07 
 &\underline{72.17} $\pm$ 0.12 &{51.43} $\pm$ 0.15   
 &\underline{50.17} $\pm$ 0.12 &33.89 $\pm$ 0.08 \cr
    MSGIFSR &50.15 $\pm$ 0.12 &27.10 $\pm$ 0.08	
    &71.71 $\pm$ 0.09 &61.80 $\pm$ 0.07	
    &49.69 $\pm$ 0.14 &36.71 $\pm$ 0.11 \cr
    MGS &\underline{52.28} $\pm$ 0.17 &\underline{35.16} $\pm$ 0.11	
    &71.87 $\pm$ 0.14 &62.53 $\pm$ 0.11	
    &50.02 $\pm$ 0.12 &37.99 $\pm$ 0.09\cr
    DGNN & 49.88 $\pm$ 0.14  & 34.17 $\pm$ 0.09	
    &71.29 $\pm$ 0.13 &62.85 $\pm$ 0.14	
    &48.45 $\pm$ 0.09 &\underline{38.10} $\pm$ 0.10 \cr
    \midrule  
    CoHHN           &54.50 $\pm$ 0.16 &35.42 $\pm$ 0.12	
    &72.87 $\pm$ 0.14 &63.35 $\pm$ 0.12	
    &51.43 $\pm$ 0.15 &38.81 $\pm$ 0.11 \cr
	{\bf \baby}     &{$\bf 55.28^*$}$\pm$ 0.12 &{$\bf 35.94^*$}$\pm$ 0.08
	               &{$\bf 73.66^*$}$\pm$ 0.13 &{$\bf 63.69^*$}$\pm$ 0.09
	               &{$\bf 52.66^*$}$\pm$ 0.11 &{$\bf 39.02^*$}$\pm$ 0.07\cr
	\bottomrule
    \end{tabular}
    \label{performance}
    \caption*{The results (\%) produced by the best baseline and the best performer in each column are underlined and boldfaced respectively. Statistical significance of pairwise differences of \baby against the best baseline (*) is determined by the t-test ($p < 0.01$).}
\end{table} 
Table~\ref{performance} presents the overall performance of all methods on three datasets, where we can get following insights:

Firstly, baselines' performance exhibits considerable discrepancy across distinct datasets. For example, MGS achieves best results among baselines in Cosmetics, while its performance in Grocery and Toys is unsatisfactory. We argue that the inconsistency of baselines' performance comes from that all baselines only focus on modeling user interest preference. All of them ignore that the price preference depending on various information is also critical when users make purchase decisions. In fact, users focus on different preferences on different datasets, \eg users possess different price sensitivity on different context. Thus, constrained by considering only single kind user preference, baselines are unable to achieve good performance across various context. 

Secondly, drawing support from attention mechanism to distinguish the importance of items within sessions, NARM and BERT4Rec perform much better than GRU4Rec. Nevertheless, the structure of BERT4Rec does not achieve amazing performance in the task of SBR as it has done in the field of Natural Language Processing (NLP). We speculate the reason for this is that the BERT structure is good at capturing distant transition patterns among tokens (\ie items in SBR) in a sequence. As suggested in Table~\ref{statistics}, the session usually contains a few items, \ie the average number of items existing in a session is 5.99, 2.38 and 2.48 on Cosmetics, Grocery and Toys respectively. Therefore, BERT4Rec can not unleash its strength in such short sequences, leading to its unsurprising performance.

Thirdly, SR-GNN and LESSR obtain remarkable results by utilizing GNN to capture pair-wise relations between adjacent items. Especially in the terms of MRR on Grocery and Toys, SR-GNN achieves strong performance among baselines. However, unable to model high-order relations existing among items, the overall performance of SR-GNN and LESSR is not optimal. Exploring internal and external connectivity of sessions, COTREC achieves competitive performance in terms of Precision metric. It builds graphs from distinct views, \ie item view and session view, to augment session data. The introduction of two views enables COTREC to model beyond pair-wise patterns within sessions, which enhances its performance. Similar as CORTEC, DGNN explores explicit and implicit item correlations by building distinct item graphs for complex relation modeling. In contrast, $S^2$-DHCN captures high-order relations among items via hypergraph convolutional networks. 
MGS performs admirably among baselines, especially in the Cosmetics. We argue that the introduction of auxiliary attribute information, \ie categories and brands, endows MGS with the ability to finely-grained mine user interest, leading to its impressive performance.
In addition, SR-GNN+ generally outperforms SR-GNN, which suggests that considering side information can improve SBR. Besides, the modified SR-GNN+ is still defeated by the proposed \baby which customizes model structures for user price preference understanding. It verifies the unique effectiveness of the proposed \baby on modeling user price preference in SBR.

Finally, our proposed \baby achieves consistent best performance on all metrics in all datasets. Concretely, compared with the best baselines, the relative improvement of \baby at Prec@20 and MRR@20 is around 5.74\% and 2.22\% on Cosmetics, 2.06\% and 0.92\% on Grocery and 4.96\% and 2.41\% on Toys. We believe that the superiority of \baby mainly comes from: (1) it considers the price preference when predicting user purchase behaviors; (2) it collectively emphasizes the importance of both price and interest preference on determining user choices. In \baby, a customized heterogeneous hypergraph with a novel triple-level convolution is devised to capture price and interest preference of users from heterogeneous information. Besides, a bi-preference learning schema is proposed to model user intent via deducing price and interest preference simultaneously. Based on above factors, our model is able to offer satisfactory personalized services. In addition, \baby outperforms our previous method CoHHN. It demonstrates that (1) it is rationale to incorporate item brands to model user price preference; (2) the triple-level convolution contributes to obtaining expressive embeddings; (3) the bi-preference learning schema is able to accurately reveal user purchase intent. 

\subsection{The Effect of Price Information (RQ2)}\label{sec:price}

\begin{table}[t]
\tabcolsep 0.05in 
  \centering
    \caption{The Effect of Price Information.}
    \renewcommand{\arraystretch}{1.1}
    \begin{tabular}{c cc cc cc}  
    \toprule  
    \multirow{2}*{methods}& 
    \multicolumn{2}{c}{Cosmetics}&\multicolumn{2}{c}{Grocery}&\multicolumn{2}{c}{Toys}\cr  
    \cmidrule(lr){2-3} \cmidrule(lr){4-5} \cmidrule(lr){6-7}
    &Prec@20&MRR@20 &Prec@20&MRR@20 &Prec@20&MRR@20\cr  
    \midrule  
    NARM    &48.52 &34.50   &71.54 &62.81    &48.42 &37.81 \cr
    NARM+p    &48.86&34.69   &71.79&62.95    &48.28&37.93 \cr
	SR-GNN  &50.75 &34.78   &71.68 &63.03    &49.48 &38.05 \cr
    SR-GNN+p  &50.86&34.97   &71.84&62.97    &49.63&38.12 \cr
    \midrule
    MGS     &52.28&35.16	&71.87&62.53	&50.02&37.99\cr
    CoHHN   &54.50&35.42    &72.87&63.35    &51.43&38.81\cr
    \baby-p     &54.38&35.12	&72.89&62.61	&51.79&37.65\cr
    \baby-pp    &54.85&35.68	&73.36&62.80	&52.25&38.03\cr
	{\bf \baby} &{$\bf 55.28^*$}&{$\bf 35.94^*$} 
	            &{$\bf 73.66^*$}&{$\bf 63.69^*$}
	            &{$\bf 52.66^*$}&{$\bf 39.02^*$}\cr
	\bottomrule
    \end{tabular}

    \label{priceEffect}
\end{table}

The main contribution of our proposed \baby is that it considers price information in SBR and regards the price preference as important as interest preference in determining user purchase behaviors. We remove the price information in the \baby to build a variant named \baby-p. Moreover, we only retain the price information to learn the node embeddings in the triple-level convolution, while not deliberately mining user price preference subsequently. We name this variant as \baby-pp. In addition, to further investigate the utility of price information in SBR, we incorporate item price into two representative baselines NARM and SR-GNN to formulate NARM+p and SR-GNN+p. Specifically, NARM+p and SR-GNN+p view the addition of item ID and price embeddings as inputs.

From Table~\ref{priceEffect}, we can observe that: (1) The \baby-pp achieves better performance than \baby-p in all cases. By utilizing item price information for updating item embeddings, the \baby-pp is able to perceive user price sensitivity, which makes it perform better than \baby-p. It suggests that considering price contributes to understanding user intent, thus increasing the recommendation accuracy. It is also in accord with reality that the price is an important factor for users to consider when shopping. (2) The \baby outperforms \baby-pp in a large margin, which indicates that we should view the price as a critical factor instead of accessory one when modeling user behaviors. It also verifies our point of view that the user price preference plays an significant role in determining user choices and we should view the price preference  as important as interest one to provide satisfactory personalized services. (3) Generally, the baseline methods considering item price outperform its counterparts without, \ie NARM+p vs NARM and SR-GNN+p vs SR-GNN. It highlights once more that taking the price information into account can improve SBR.

\subsection{The Effect of Price Discretization (RQ2)}\label{sec:priceDis}
\begin{figure}[t]
  \centering
  \includegraphics[width=0.9\linewidth]{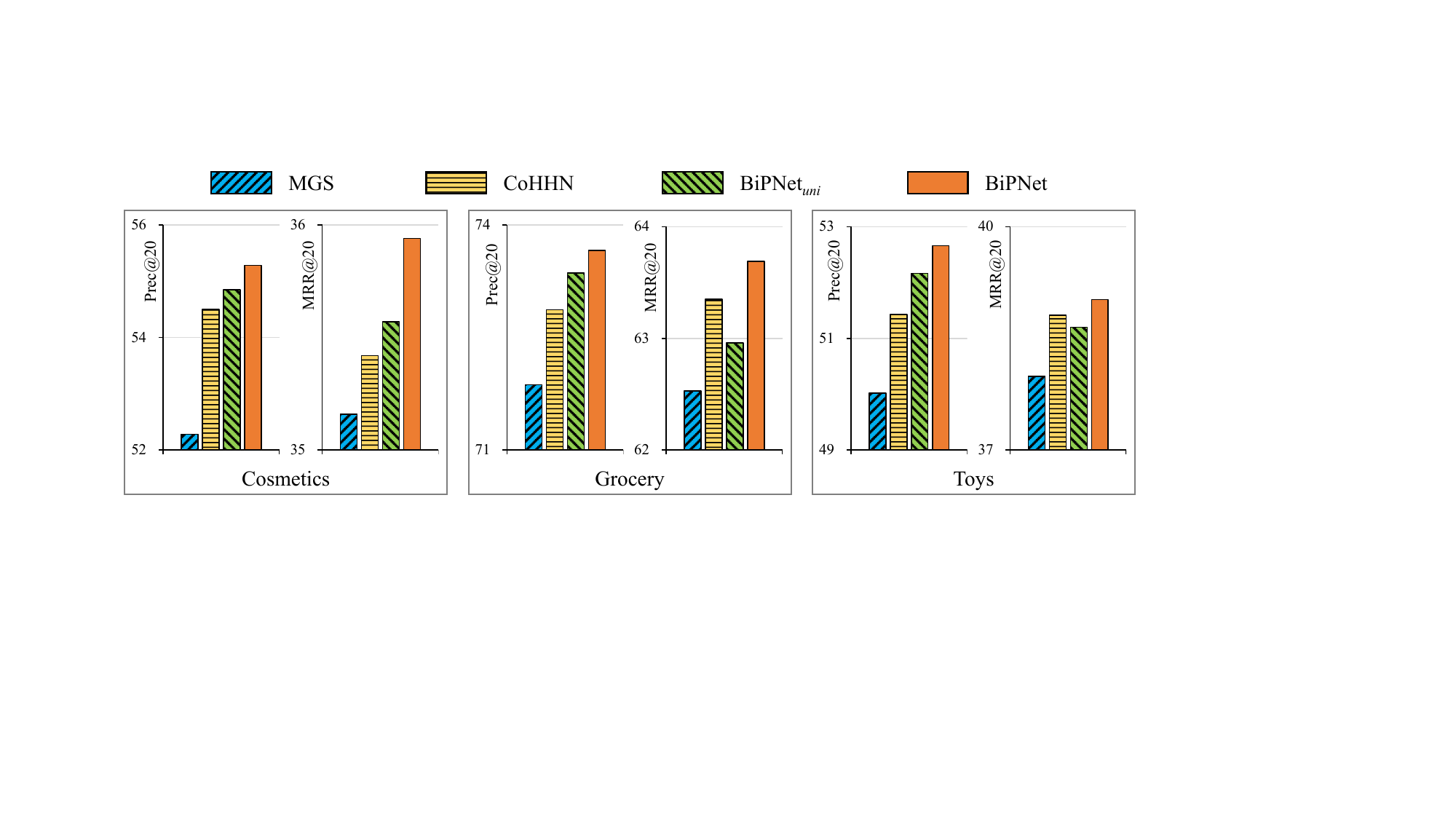}
  \caption{The effect of the price discretization.}\label{logistics}
\end{figure}

As discussed in Section~\ref{sec:pricedis}, to compare the degree of price across various categories, we divide absolute price into price levels. Different from common practice where item price levels are simply defined by uniform quantization, we find that the price distribution tends to be logistic distribution and perform price discretization via equally partitioning the logistic probability distribution. The \babyuni$_{uni}$ obtains the price levels by uniform quantization like in~\cite{Zheng@ICDE2020}, where $ p_i = \lfloor \frac{x_p - min}{max - min} \times \rho \rfloor $. 

\fig ~\ref{logistics} presents the results of MGS, CoHHN, \babyuni$_{uni}$ and \baby in terms of Prec@20 and MRR@20 on all datasets. \baby achieves better performance than \babyuni$_{uni}$, which indicates the superiority of our proposed method for price discretization. We conjecture the reasons for the good performance are: (1) the price distribution is more compatible with logistic distribution rather than uniform one; (2) equal probability interval makes the price reasonably arranged in each price level and enables the model to finely perceive price preference of users. In addition, \babyuni$_{uni}$ outperforms MGS by a large margin, which proves the benefits of introducing price into SBR again.

\subsection{The Influence of Heterogeneous Information (RQ2)}\label{sec:heter}

\begin{table}[t]
\tabcolsep 0.05in 
  \centering
    \caption{The Influence of Heterogeneous Information.}
    \renewcommand{\arraystretch}{1.1}
    \begin{tabular}{c cc cc cc}  
    \toprule  
    \multirow{2}*{methods}& 
    \multicolumn{2}{c}{Cosmetics}&\multicolumn{2}{c}{Grocery}&\multicolumn{2}{c}{Toys}\cr  
    \cmidrule(lr){2-3} \cmidrule(lr){4-5} \cmidrule(lr){6-7}
    &Prec@20&MRR@20 &Prec@20&MRR@20 &Prec@20&MRR@20\cr  
    \midrule  
    {(a) w/o p}     &54.38&35.12	&72.89&62.61	&51.79&37.65\cr
    {(b) w/o c}     &54.43&35.45	&73.21&63.05	&52.18&38.72\cr  
    {(c) w/o b}     &54.74&35.67	&72.93&63.07	&51.90&38.80\cr
    {(d) w/o pc}    &53.01&35.09	&73.16&62.68	&52.19&37.66\cr
    {(e) w/o pb}    &54.54&35.56	&73.03&62.55	&51.96&37.45\cr
    {(f) w/o cb}    &54.27&35.41	&72.18&62.81	&51.11&38.63\cr
    {(g) w/o pcb}   &52.15&32.67	&71.87&61.63	&50.94&36.84\cr
	{\bf \baby} &{$\bf 55.28^*$}&{$\bf 35.94^*$} 
	            &{$\bf 73.66^*$}&{$\bf 63.69^*$}
	            &{$\bf 52.66^*$}&{$\bf 39.02^*$}\cr
	\bottomrule
    \end{tabular}

    \label{heterogeneousInfor}
\end{table}

The user price preference varies greatly according to item features, \eg category and brand. Therefore, besides item price, we introduce extra information into our \baby to facilitate price preference modeling. In this part, we zoom into each type information to examine its influence on \baby's performance. The ablation results are shown in Table~\ref{heterogeneousInfor}, where  'p', 'c' and 'b' mean price, category and brand information respectively. We can get following insights from Table~\ref{heterogeneousInfor}: 

(1) The variant (g) without incorporating any extra information obtains the worst performance. It proves that considering extra features in SBR does contribute to revealing user intent, increasing the prediction accuracy;
(2) Except the \baby, there is no variants consistently performing well on all cases. It indicates that the user behaviors present complexity and are influenced by multiple item features. Going further, each type of information we consider in the model has unique impact on user decisions in different context;
(3) In general, the variants incorporating price information perform better than the variants do not, \eg variants (b, c) against (a). Especially, variant (f) which only considers item price achieves overwhelming superiority over variant (g) which does not incorporate any side information. It demonstrates again that price is indeed a crucial factor in determining user behaviors and taking it into account brings significant improvements for SBR;
(4) The variant only containing price while ignoring other factors performs worse than variants considering price and related features, \eg variant (f) is defeated by variant (b, c). It verifies our hypothesis that the user price preference is related to different item features. Thus, only incorporating price information without considering related features can not accurately extract price preference of users;
(5) \baby achieves best performance in all cases, which demonstrates that the proposed model can handle the heterogeneous information in SBR, leading to good understanding of user behaviors.


\subsection{The Effect of Triple-level Convolution (RQ2)}\label{sec:effecttripple}

\begin{figure}[t]
  \centering
  \includegraphics[width=0.9\linewidth]{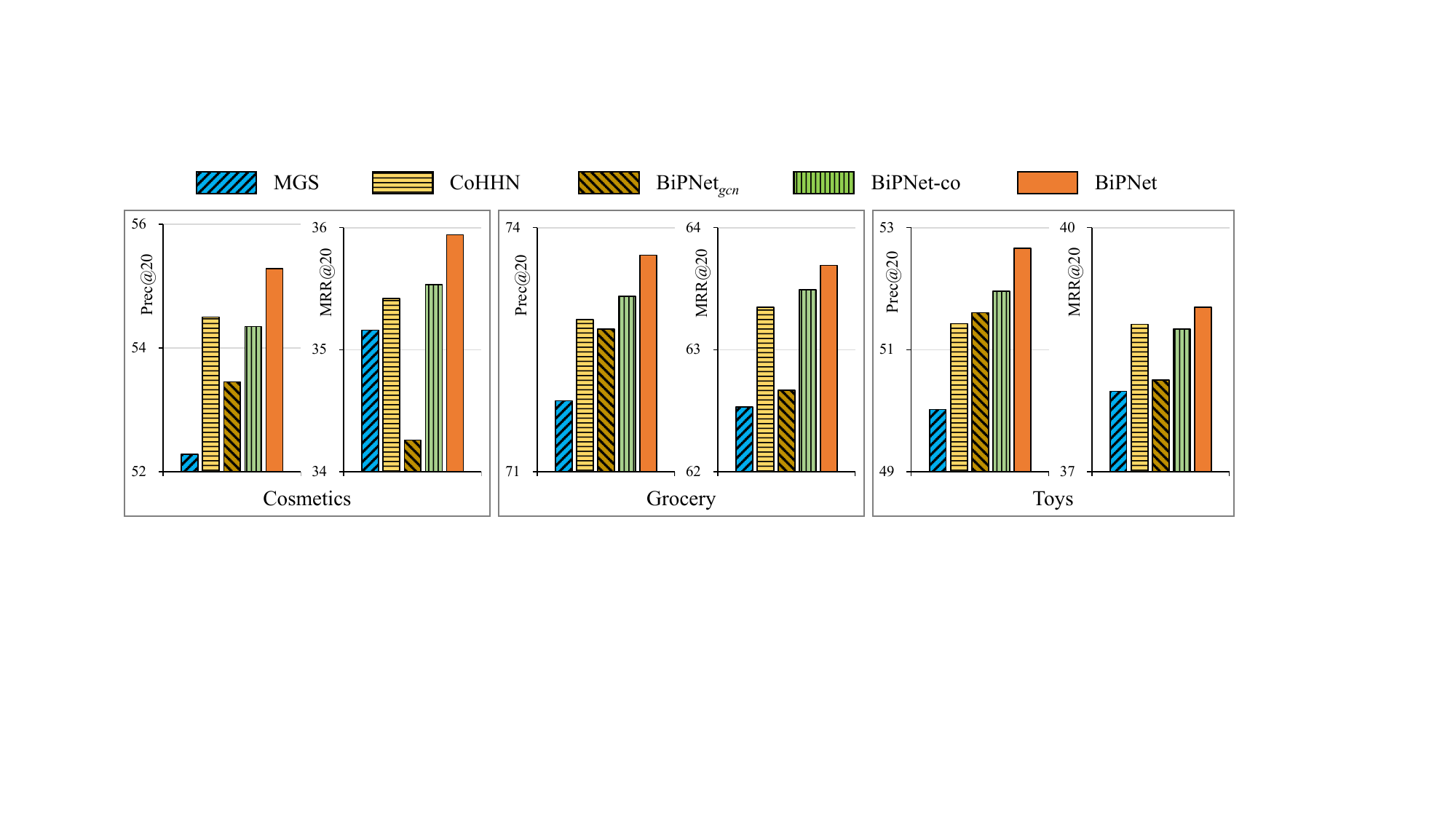}
  \caption{The effect of triple-level convolution.}\label{tripleLevel}
\end{figure}

Considering that there are mainly three kinds of relations among heterogeneous nodes, \ie co-occurrence relations, intra-type relations and inter-type relations, we propose a novel triple-level convolution to aggregate rich information in the heterogeneous hypergraph for node representation learning. To demonstrate the effectiveness of this design choice, we replace the triple-level convolution with commonly used GCN like in~\cite{Zheng@ICDE2020} to formulate the variant \babyuni $_{gcn}$. That is, for a target node, \babyuni $_{gcn}$ obtains its embedding by linearly aggregating embeddings of its adjacent nodes without regard to distinct relations among them. Moreover, compared with CoHHN, \baby further enhances the node representation learning by performing co-occurrence convolution. To examine its utility, we discard co-occurrence convolution from \baby and obtain a variant \baby-co. 

As presented in \fig~\ref{tripleLevel}, \baby performs much better than \babyuni $_{gcn}$ in both metrics on all datasets, which shows the superiority of the proposed triple-level convolution in representation learning under such complex heterogeneous situation. We believe its superiority comes from that the triple-level convolution finely distinguishes distinct relations existing among various nodes and designs customized method to extract meaningful information for every relation. It endows the \baby with the capacity to obtain accurate node embeddings from rich heterogeneous information, thus clearly revealing user intent. Besides, \babyuni $_{gcn}$ considering item price, categories and brands generally performs better than MGS incorporating item categories and brands (except on MRR@20 in Cosmetics), which demonstrates the significance of price in influencing user behaviors in SBR again. 
In addition, \baby achieves better performance than \baby-co, which verifies the effectiveness of co-occurrence convolution in \baby. It is intuitive since that nodes with co-occurrence relations possess similar semantics which can contribute to learning meaningful node embeddings. 

\subsection{The Effect of Bi-preference Learning Schema (RQ2)}\label{sec:bipre}
\begin{table}[t]
\tabcolsep 0.1in 
  \centering
    \caption{The Effect of Bi-preference Learning Schema.}
    \renewcommand{\arraystretch}{1.1}
    \begin{tabular}{c cc cc cc}  
    \toprule  
    \multirow{2}*{methods}& 
    \multicolumn{2}{c}{Cosmetics}&\multicolumn{2}{c}{Grocery}&\multicolumn{2}{c}{Toys}\cr  
    \cmidrule(lr){2-3} \cmidrule(lr){4-5} \cmidrule(lr){6-7}
    &Prec@20&MRR@20 &Prec@20&MRR@20 &Prec@20&MRR@20\cr  
    \midrule  
    MGS     &52.28&35.16	&71.87&62.53	&50.02&37.99\cr
    CoHHN     &54.50&35.42   &72.87&63.35    &51.43&38.81\cr
    \baby-BiP   &54.76&35.43	&73.21&62.80	&51.86&38.41\cr
    \babyuni$_{CE}$   &54.99&35.74 &73.41&62.52 &52.38&38.53\cr
	{\bf \baby} &{$\bf 55.28^*$}&{$\bf 35.94^*$} 
	            &{$\bf 73.66^*$}&{$\bf 63.69^*$}
	            &{$\bf 52.66^*$}&{$\bf 39.02^*$}\cr
	\bottomrule
    \end{tabular}

    \label{bip}
\end{table} 

As stated in former sections, a user often makes price-interest trade-off when she shops online. Obviously, there are complex mutual relations between user price preference and interest preference. Both price and interest preference are indispensable in SBR and they collectively determine user choices. To accurately capture these two preferences, we propose a bi-preference learning schema under the multi-task learning architecture. In order to validate the effectiveness of the proposed schema, we remove it from the model to build a variant \baby-BiP. The \baby-BiP directly applies the price $\mathbf{\hat{u}}_\mathsf{p}$ and interest preference $\mathbf{\hat{u}}_\mathsf{I}$ into Equation (35) to formulate the recommendation list. Besides, we replace the multi-task learning paradigm with commonly used cross-entropy loss~\cite{NARM,SR-GNN,MGS,CoHHN} to formulate \babyuni$_{CE}$. That is, \babyuni$_{CE}$ applies single loss based on Equation (35) for model training while failing to incorporate extra signals to collectively deduce user price and interest preference. 

As shown in Table~\ref{bip}, \baby outperforms \baby-BiP in terms of Prec@20 and MRR@20 in all datasets, which indicates the effectiveness of the proposed bi-preference learning schema for the task. It also suggests that there exists complex mutual relations in price and interest preference. The bi-preference learning schema can explore the mutual relations between price and interest preference, leading to \baby's good performance. 
Moreover, \babyuni$_{CE}$ is defeated by \baby, which demonstrates the rationale and necessity of the multi-task learning paradigm in our settings. With the help of the multi-task learning architecture, the proposed \baby is able to accurately deduce user price and interest preference, contributing to its prediction for user behaviors.
In addition, both \baby-BiP and \babyuni$_{CE}$ achieve much better performance than MGS, especially in terms of Prec@20. We believe that the good performance comes from their modeling for price preference, which proves the significance of taking item price into account once more in SBR.

\subsection{Performance under Different Price Levels (RQ3)}\label{sec:priceLevel}

\begin{figure}[t]
  \centering
  \includegraphics[width=0.9\linewidth]{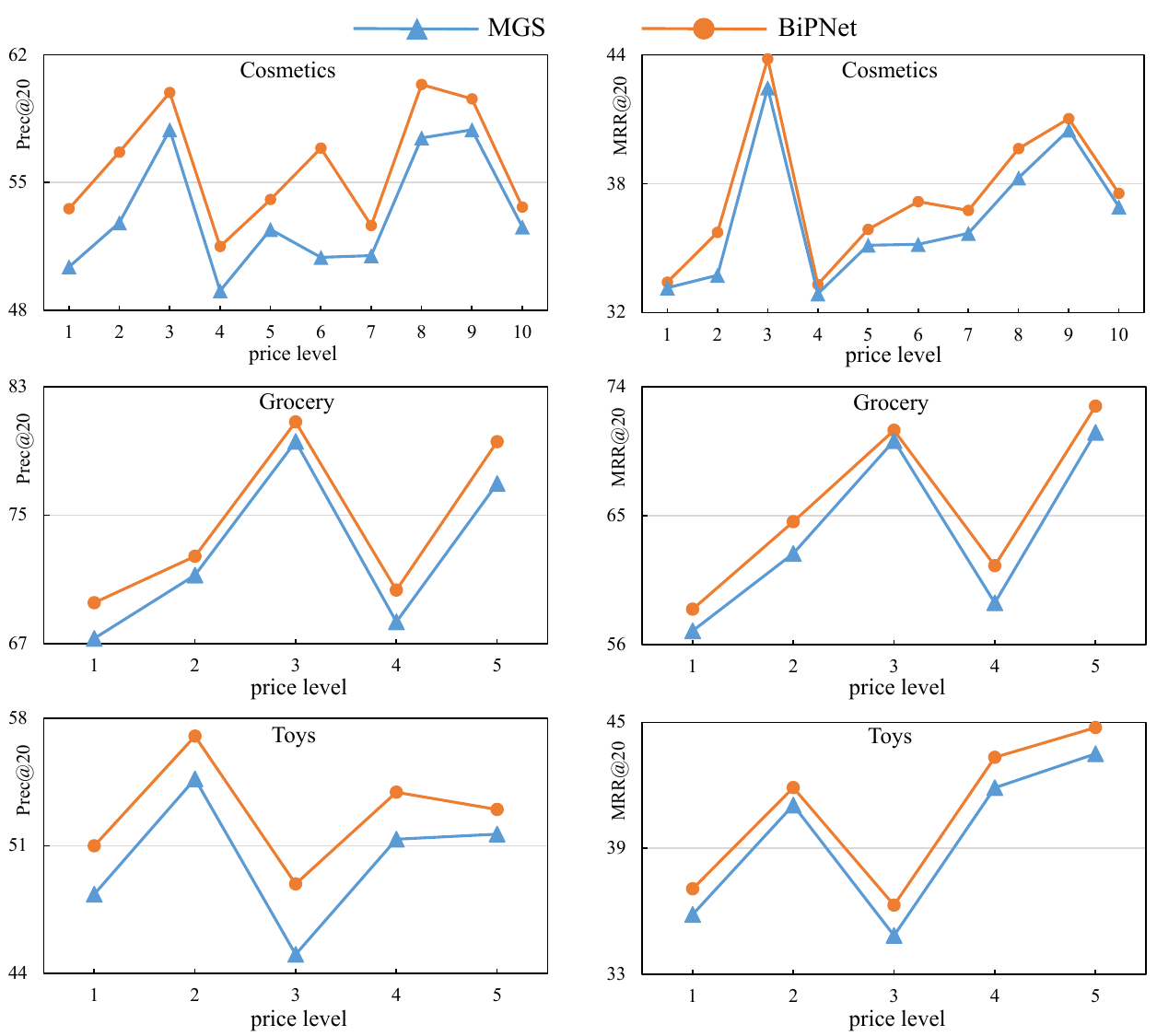}
  \caption{Performance under different price levels.}\label{priceLevel}
\end{figure}

The former sections have indicated that both price and interest preference contribute to predicting user purchase behaviors. Zooming in each preference, nevertheless, users present great behavioral discrepancy. With possessing distinct tastes, there are high personalization in interest preference for different users. For example, some users like T-shirts with round neck, while some others may prefer polo shirt. As to price preference, in contrast, most users tend to prefer items with low price. This is straightforward to understand since that people always want to spend as little money as possible to get desired items. Thus, we could improve the overall performance by blindly providing users with cheap items. However, such a move will cause decline of business revenue given that expensive items always offer huge profits. Therefore, we conduct experiments under different price levels to examine the performance of the proposed \baby and competitive baseline MGS. \fig~\ref{priceLevel} plots their performance patterns on all datasets in terms of Prec@20 and MRR@20, where we can get following observations:

On the one hand, \baby outperforms MGS on all cases in all datasets. It indicates that our \baby can understand user needs in different price levels and provide accurate recommendation accordingly. In other words, \baby can not only improve user satisfaction, but also increase the income of business. We believe that this attribute of the \baby can promote the benign development of e-commerce.  
On the other hand, models' performance patterns show different trends in distinct datasets. More specifically, in Cosmetics and Toys, models perform better in low (\eg 3 in Cosmetics and 2 in Toys) and high (\eg 8, 9 in Cosmetics and 4, 5 in Toys) price than in medium one (\eg 4, 5, 6, 7 in Cosmetics and 3 in Toys). Instead, in Grocery, they achieve good performance in medium price (\eg  3), while perform badly in low (\eg 1, 2) and high price (\eg 4). We speculate that the discrepancy comes from distinct behavior characteristics in different context (datasets). For instance, in Cosmetics, users usually either choose cheap items for daily use or buy expensive ones for occasional cases. While items with medium price are rarely noticed. As to Grocery, the quality of cheap food is worrying. Also, it is unnecessary to buy expensive ones in most cases. Thus, items with medium price are commonly popular. Obviously, considering item price in e-commerce can enable the business to detect user fine-grained  behavior characteristics and offer satisfactory personalized services accordingly.

\subsection{Impact of the Session Length (RQ4)}\label{sec:sessionlength}
\begin{figure}[t]
  \centering
  \includegraphics[width=0.9\linewidth]{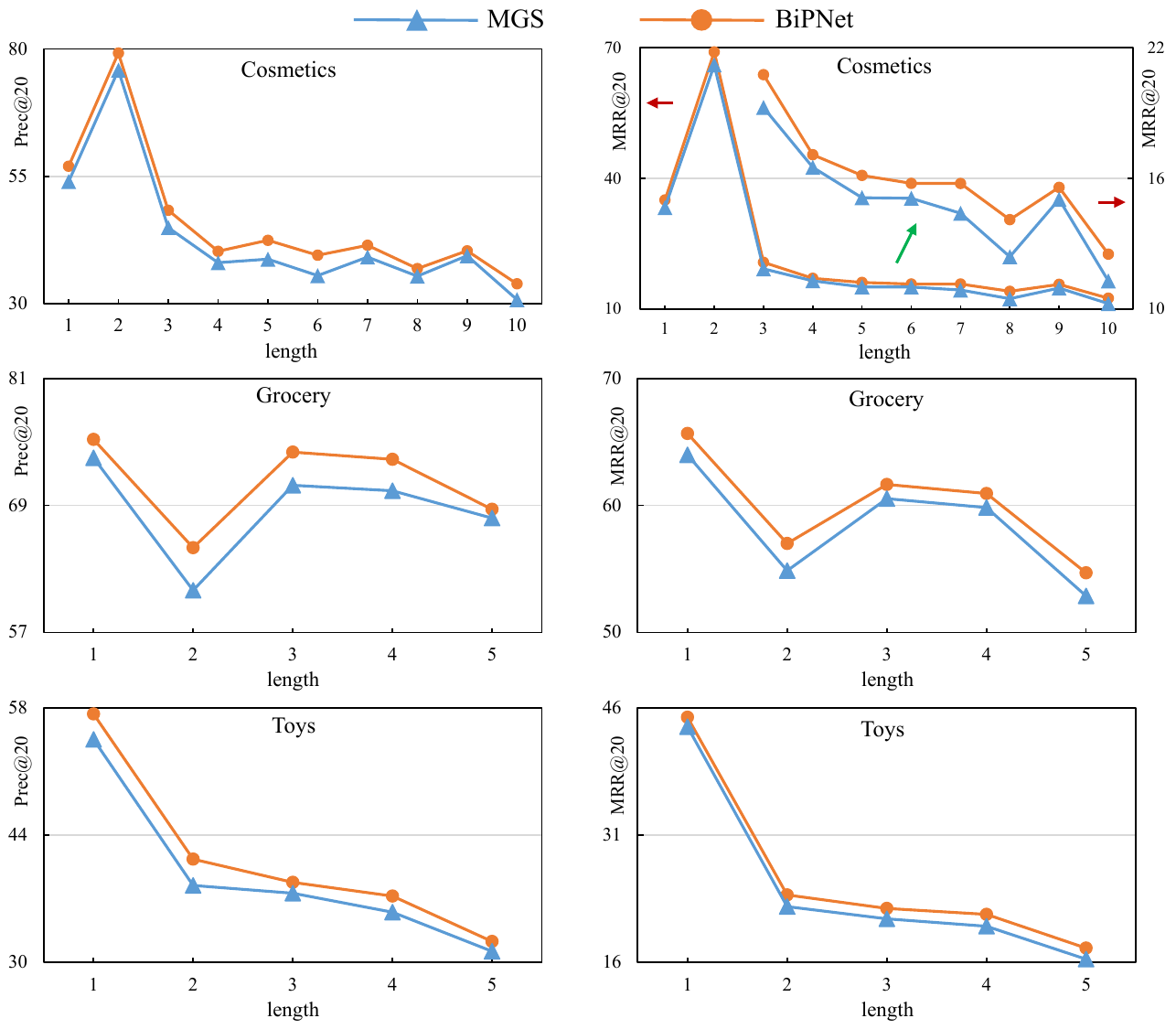}
  \caption{Impact of the session length.}\label{sessionlength}
\end{figure}

The session length is a key factor that influences model performance in SBR, since that it signifies how much information the model can rely on to capture user preference. In this part, therefore, we explore the performance of \baby and representative baseline MGS under different session length in all datasets. Based on average session length of the dataset (as shown in Table~\ref{statistics}), we choose different ranges to display models' performance patterns in three datasets, \ie [1, 10] for Cosmetics and [1, 5] for Grocery/Toys. The \baby's and MGS's performance curves in terms of Prec@20 and MRR@20 are presented in Figure~\ref{sessionlength}. Note that, due to significant performance gap existing on MRR@20 in Cosmetics, we set two unique scales in y-axis to highlight the performance patterns, \ie we can refer to the right y-axis to see the performance of length 3-10 for clarity. The following insights can be concluded from Figure~\ref{sessionlength}:

(1) \baby outperforms MGS in all cases, which demonstrates the effectiveness of \baby in the task of SBR. Compared with MGS, the proposed \baby not only considers user interest preference but also emphasizes the significance of user price preference hidden in  various item features like price, categories and brands. We contend that the consistently strong performance of \baby on various session length comes from its careful consideration for user price preference.
(2) Generally speaking, as the session length increases, the performance of both \baby and MGS deteriorates. That is, models perform unsatisfactorily under long sessions. As suggested in previous works, in long sessions, there may exist much more noisy clicks~\cite{DIDN} and multiple user intents~\cite{Zhang@WSDM2023}. As a result, it is challenging for these models to accurately predict user behaviors under such a complex situation.
(3) Overall, compared with long sessions, \baby achieves larger improvements over MGS in short ones. In sessions with a few items (\eg session length is 1 or 2), there is a little information available for models to capture user intent. In another word, the models have a serious problem with data sparsity. Fortunately, the incorporation of item side information in \baby enriches the data, alleviating data sparsity issue to some extent. It also reminds us the rationale of modeling various user preferences in SBR.

\subsection{Hyper-parameter Study (RQ5)}\label{sec:hyper}

\begin{figure}[t]
\centering
    \subfigure[{Influence of the number of price levels $\rho$.}]{
        \includegraphics[width=0.9\linewidth]{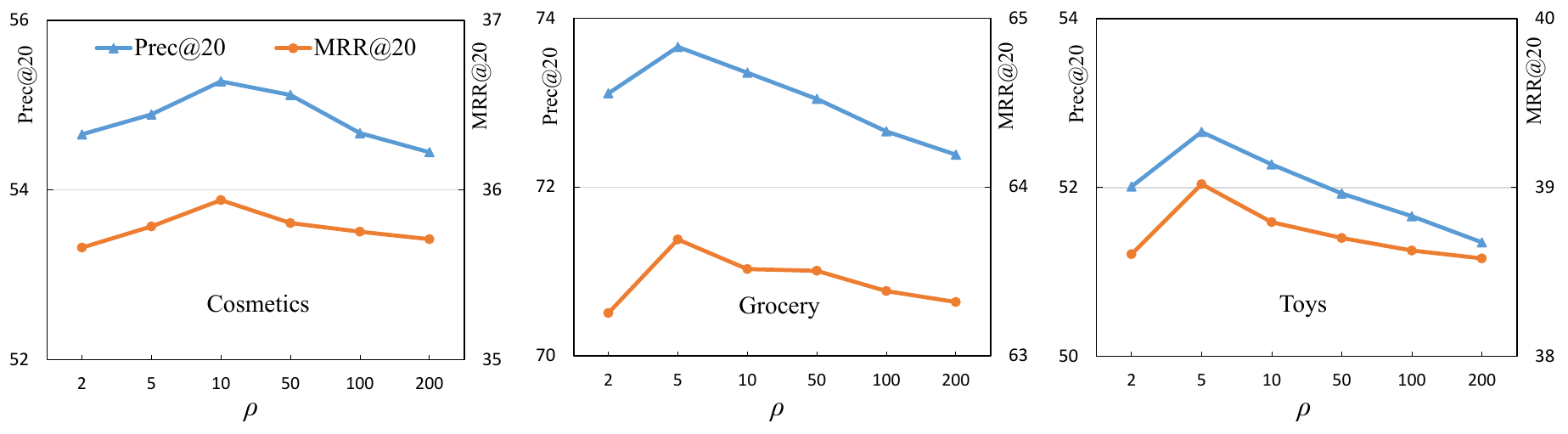}
    }
    \subfigure[{Influence of the repeating number for the triple-level convolution $r$.}]{
        \includegraphics[width=0.9\linewidth]{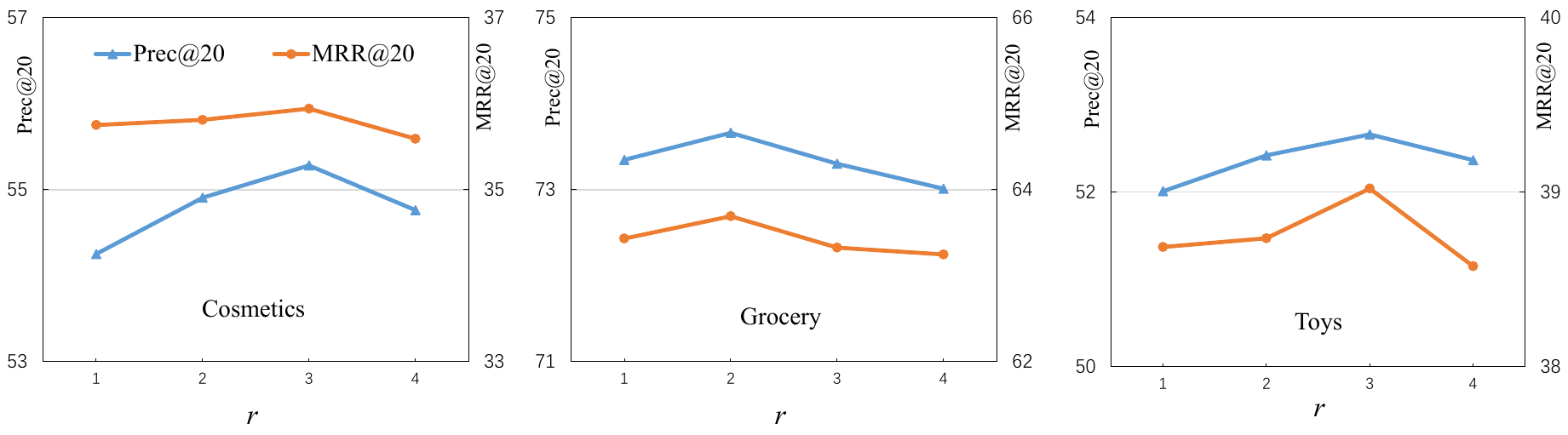}
    }
    \subfigure[{Influence of the number of self-attention heads $h$.}]{
        \includegraphics[width=0.9\linewidth]{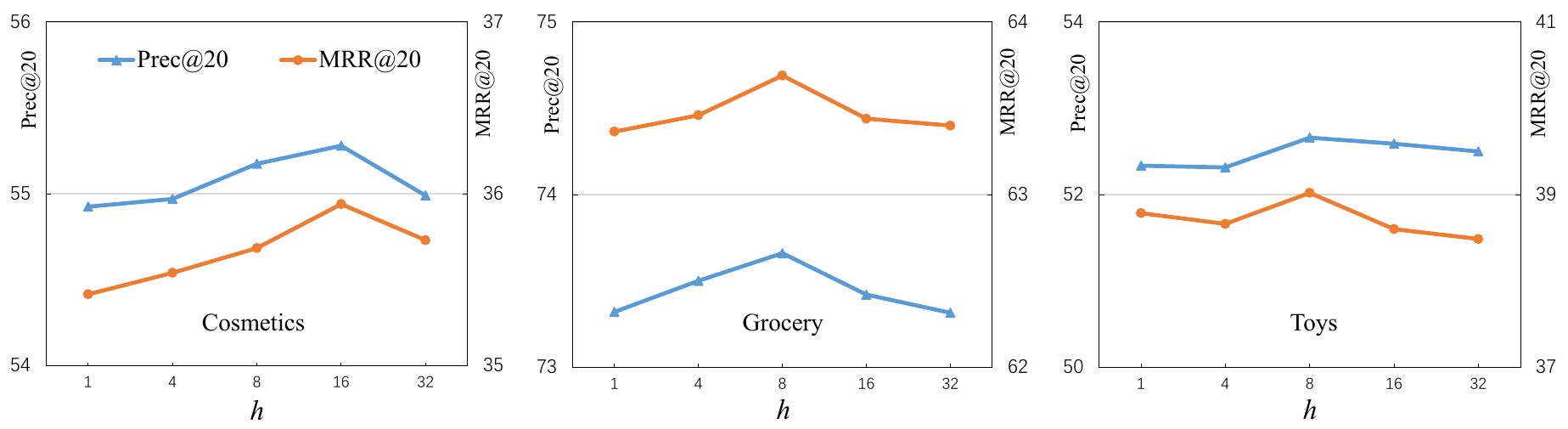}
    }
    \caption{\babyuni's performance under different hyper-parameters}\label{hyperparameter}
\end{figure}

In this part, we investigate the influence of three main hyper-parameters, \ie the number of price levels $\rho$, the repeating number of triple-level convolution $r$ and the number of self-attention heads $h$, on our proposed \baby.

\paratitle{The number of price levels $\rho$} controls the fineness of price levels. The larger the  $\rho$, the finer the granularity of price discretization, where the price sensitivity of users is assumed higher.  
As can be seen from \fig~\ref{hyperparameter}(a), if we define $\rho$ too small, \ie 2, which signifies that an item price is coarsely viewed as either expensive or cheap, \baby can not perceive different price preferences, leading to invalid price modeling and unsatisfactory predictions. In contrast, when the $\rho$ is too large like 200, the similar price is allocated into different price levels, leading to fragmented price levels. In such a case, the difference among price levels is not evident, which is also detrimental to model performance. Moreover, in the view of absolute value, the performance under different $\rho$ is stable. We believe that the stability comes from the method we developed to form price levels. Equally partitioning the price probability distribution, our method effectively allocates absolute price into each level and balances the training data, which helps the model accurately capture various price preferences. In addition, \baby achieves best results under different $\rho$ in different datasets, \ie 10 for Cosmetics and 5 for Grocery/Toys. We guess that users present different price sensitivity under different context (datasets), leading to different optimal values for $\rho$.

\paratitle{The repeating number of triple-level convolution $r$} determines how much information a node incorporates from its adjacent nodes. Obviously, every time we conduct the operation of triple-level convolution, the target node may collect some information from its neighbors. As shown in \fig~\ref{hyperparameter}(b), with $r$ increasing, the \baby's performance improves first because that nodes can obtain more useful information from others, and then drops due to over-smooth issues. We choose different $r$ for different datasets, \ie 3 for Cosmetics/Toys and 2 for Grocery, according to experimental results. We speculate that there are fewer items in Grocery than Cosmetics and Toys (as shown in Table~\ref{statistics}), which enables the proposed \baby to extract semantics of nodes with fewer triple-level convolution iterations in Grocery.

\paratitle{The number of self-attention heads $h$} determines the granularity of partitioning subspaces when handling price sequences, where the model attends to information from different aspects. As presented in \fig~\ref{hyperparameter}(c), too little (\eg 1) or too much (\eg 32) heads can have a negative impact on \baby's performance. We argue that few heads are unable to capture complex price preference, while many heads make the semantics excessively dispersion, both leading to declining performance. 
Moreover, referring to \fig~\ref{hyperparameter}(c) and Table~\ref{statistics}, we observe that dataset with long sessions (Cosmetics) benefits from a large $h$ (16) while datasets with short sessions (Grocery and Toys) prefer a small $h$ (8). It is in line with the fact that self-attention with large number of heads is good at capturing distant dependencies among sequences. 

\subsection{Complexity Analysis (RQ6)}\label{sec:complexity}
In this part, we analysis the time complexity of the proposed \baby. The most time-consuming part of \baby is the representation learning of triple-level convolution in the proposed heterogeneous hypergraph. We can detail its time complexity as $O$($n \times r \times $ ($\bar{N}_{id} + \bar{N}_{p} + \bar{N}_{c} + \bar{N}_{b}$) ). Specifically, $n$ is the total number of items, $r$ is the iterative number of triple-level convolution, and $\bar{N}_{id}$, $\bar{N}_{p}$, $\bar{N}_{c}$ as well as $\bar{N}_{b}$ are the average number of adjacent nodes with type ID, price, category and brand for items. Referring to Table~\ref{statistics}, the number of item price, category and brand is relatively small, leading to their limited contribution on increasing model complexity. For example, there are totally 5 price levels in Grocery, \ie the $\bar{N}_{p}$ is no more than 5. That is, although we incorporate various kinds of information to depict user behaviors, the model is increased on a modest scale. Besides, we can also reduce the number of side features, like merging similar categories or pruning some niche brands, to improve model efficiency. We believe that such merits of \baby contribute to its application in real scenarios.

\section{Conclusion and Future Work}\label{sec:conclusion}
The user purchase behaviors are determined by not only interest preference but also price preference in real-world scenario. Unfortunately, the existing methods for session-based recommendation only focus on user interest preference. Therefore, in this work, we propose a novel approach Bi-Preference Learning Heterogeneous Hypergraph Networks (\baby) to capture user interest and price preference simultaneously via exploring various kinds of information for session-based recommendation. As the pioneer work considering item price in SBR, \baby identifies two challenges that get in the way of utilizing price information: (1) price preference is hidden behind rich heterogeneous information; (2) price and interest preference collectively determine user choice. To copy with the first challenge, in \baby, we construct a novel heterogeneous hypergraph which encodes rich information about user price preference including item ID, item price, item category and item brand. Based on the heterogeneous hypergraph, a customized triple-level convolution is innovatively devised to handle heterogeneous information for interest and price preferences learning. 
As to the second challenge, we propose a bi-preference learning schema under the multi-task learning architecture. With exploring the mutual relations between interest preference and price preference, the bi-preference learning schema is able to model these two preferences simultaneously and accurately capture user purchase intents. 
We conduct extensive experiments on three real-world datasets and the results demonstrate the superiority of the proposed \baby over representative baselines. Further study also supports that item price is of great significance to determining user behaviors in session-based recommendation.

In the future, first of all, we plan to introduce more available information like item ratings, item images and user reviews to further explore item features and capture user complex preferences.
Secondly, it is also a promising direction that distinguishing the importance of distinct user preferences under difference context for further improving prediction accuracy. For instance, automatically assigning different weights for distinct kinds of losses based on their utility. It can endow the model with the potential to identify the most influential factors on user behaviors. 
Last but not least, the proposed heterogeneous hypergraph with triple-level convolution can be easily extended to other tasks, where it is necessary to model various kinds of information with complex relations.

\bibliographystyle{ACM-Reference-Format}
\bibliography{ref}

\end{document}